\documentclass[manuscript,screen,nonacm,acmsmall]{acmart}

\usepackage{adjustbox}
\usepackage[linesnumbered]{algorithm2e}
\usepackage{array}
\usepackage[epsilon,altpo]{backnaur}
\usepackage{bm}
\usepackage{caption}
\usepackage{environ}
\usepackage{graphicx}
\usepackage{multirow}
\usepackage{subcaption}
\usepackage{suffix}
\usepackage{tikz}
\usepackage{wrapfig}
\usepackage{xcolor}
\usepackage{xspace}
\usepackage{enumitem}

\usetikzlibrary{calc}
\usetikzlibrary{positioning}

\makeatletter
\def\myparagraph{\@ifstar\@mypara\@@mypara}
\def\@mypara#1{\vspace{1ex plus 1ex minus .1ex}\noindent\textbf{#1}}
\def\@@mypara#1{\@mypara{#1}\hspace{.3em}}
\makeatother

\newcommand{\akash}[1]{\textcolor{cyan}{#1}}
\newcommand{\akashc}[1]{\textcolor{cyan}{[Akash: #1]}}

\newcommand{\hideableoutline}[1]{%
    \akashc{Outline}
    #1
    \akashc{Text}}

\newcommand{\ag}[1]{\textcolor{magenta}{#1}}
\newcommand{\agc}[1]{\textcolor{magenta}{[Aarti: #1]}}

\newcommand{\mikec}[1]{\textcolor{orange}{[Mike: #1]}}

\newcommand{\smc}[1]{\textcolor{blue}{[Sharad: #1]}}

\newcommand{\gfc}[1]{\textcolor{green}{[Grigory: #1]}}

\NewDocumentEnvironment{ignore}{+b}{}{}

\renewcommand{\hideableoutline}[1]{}

\renewcommand{\akash}[1]{#1}
\renewcommand{\akashc}[1]{}
\renewcommand{\ag}[1]{#1}
\renewcommand{\agc}[1]{}

\renewcommand{\mikec}[1]{}

\renewcommand{\smc}[1]{}

\renewcommand{\gfc}[1]{}

\newcommand{\keywordstyle}[2][1]{%
    #2%
}

\newcommand{\declarekeyword}[3][s]{%
  \newcounter{keyworddefs#2}%
  \expandafter\newcommand\csname #2\endcsname{%
      \keywordstyle[\value{keyworddefs#2}]{#3}\xspace}%
  \expandafter\newcommand\csname #2s\endcsname{%
      \keywordstyle[\value{keyworddefs#2}]{#3#1}\xspace}%
}

\newcommand{\sys}{\textsc{Bolt}\xspace}
\newcommand{\lang}{\textsc{Hex}\xspace}

\newcommand{\stepT}{\emph{Align}\xspace}
\newcommand{\stepR}{\emph{Relate}\xspace}
\newcommand{\stepV}{\emph{Verify}\xspace}

\declarekeyword{mapping}{compiler-to-accelerator mapping}
\declarekeyword{camapping}{c2a-mapping}

\declarekeyword{aspfue}{MMIO code}
\declarekeyword{Aspfue}{MMIO code}
\declarekeyword{ASPFue}{MMIO Code}

\declarekeyword{aspf}{accelerator \lang code}
\declarekeyword{Aspf}{Hex code}
\declarekeyword{ASPF}{Hex Code}

\declarekeyword{cspf}{IR code}
\declarekeyword{Cspf}{IR code}
\declarekeyword{CSPF}{IR Code}

\newcommand{\tpp}{TPP\xspace}
\declarekeyword{tppfull}{tensor product program}

\declarekeyword{skel}{sync-skeleton}
\declarekeyword{Skel}{Sync-skeleton}
\declarekeyword{skeleton}{Sync-Skeleton}

\declarekeyword[es]{dcs}{layout-sketch}
\declarekeyword[es]{Dcs}{Layout-sketch}
\declarekeyword[es]{dcsfull}{Layout-Sketch} 

\declarekeyword{layoutmap}{layout mapping}

\declarekeyword{wellaligned}{well-aligned}

\definecolor{loopColor}{rgb}{.55,.2,.85}
\definecolor{dataColor}{rgb}{.85,.43,.2}
\definecolor{cfragColor}{rgb}{.45,.15,.75}
\definecolor{afragColor}{rgb}{.88, 0.65, 0.0}
\definecolor{explicationBox}{rgb}{.75,.15,.1}

\colorlet{csKeyword}{loopColor}
\definecolor{csMisaligned}{rgb}{.9,.2,.1}
\definecolor{csUpdated}{rgb}{.3,.55,.85}

\newcounter{cspfLoop}
\NewDocumentEnvironment{csfrag*}{O{} b}{%
    \begin{flalign*}
        #2
    \end{flalign*}%
}{}
\NewDocumentEnvironment{csfrag}{O{} b}{%
    \setcounter{cspfLoop}{0}%
    \begin{csfrag*}
        #2
    \end{csfrag*}%
    \setcounter{cspfLoop}{-1}%
}{}

\newcommand{\csLoopStyle}[5][csKeyword]{%
    \settoheight{\loopOpHeight}{map-reduce}%
    \settodepth{\loopOpDepth}{map-reduce}%
    \ifx{}{#2}%
    \else%
        \mathopen{}\mathcolor{#1}{\,^{\tiny\textbf{#2}}}\hspace{-0.2em}%
    \fi%
    \mathcolor{#1}{#3}_{#4}^{#5}
}

\newcommand{\csLoop}[4][csKeyword]{%
    \ifnum\value{cspfLoop}<0
        \csLoopStyle[#1]{}{#2}{#3}{#4}%
    \else
        \stepcounter{cspfLoop}%
        \csLoopStyle[#1]{\arabic{cspfLoop}}{#2}{#3}{#4}%
    \fi%
}

\DeclareMathOperator{\csTensor}{tensor}
\DeclareMathOperator{\csLetOp}{let}
\DeclareMathOperator{\csInOp}{in}
\DeclareMathOperator{\csAsOp}{as}
\newcommand{\csLet}[1][csKeyword]{\mathcolor{#1}{\csLetOp}}
\newcommand{\csIn}[1][csKeyword]{\mathcolor{#1}{\csInOp}}
\newcommand{\csAs}[1][csKeyword]{\mathcolor{#1}{\csAsOp}}

\newlength{\loopOpHeight}
\newlength{\loopOpDepth}
\newcommand{\loopOpRule}{%
    \rule[-\loopOpDepth]{0mm}{\loopOpDepth}%
    \rule{0mm}{\loopOpHeight}%
}
\DeclareMathOperator*{\csMapOp}{map\loopOpRule}
\DeclareMathOperator*{\csRedOp}{reduce\loopOpRule}
\DeclareMathOperator*{\csLoopHoleOp}{%
    \hspace{.1em}\boxed{\:\:{\scriptstyle?}\:\:}\loopOpRule}
\newcommand{\csMap}[3][csKeyword]{\csLoop[#1]{\csMapOp}{#2}{#3}}
\newcommand{\csRed}[3][csKeyword]{\csLoop[#1]{\csRedOp}{#2}{#3}}
\newcommand{\csIter}[3][csKeyword]{\csLoop[#1]{\csLoopHoleOp}{#2}{#3}}

\newcommand{\csExprOp}{?_{\text{expr}}}
\newcommand{\csExpr}[2][csKeyword]{\boxed{\csExprOp{\scriptstyle(#2)}}}

\newcommand{\R}{\ensuremath{\mathbb{R}}}

\DeclareMathOperator{\tAdpf}{AFloat}

\definecolor{asKeywordColor}{HTML}{6d8da6}
\newcommand{\asKword}[1]{\ensuremath{\mathcolor{asKeywordColor}{\bm{#1}}}\xspace}
\newcommand{\asLoopKword}[1]{\ensuremath{\mathcolor{loopColor}{\bm{#1}}}\xspace}
\newcommand{\asDataKword}[1]{\ensuremath{\mathcolor{dataColor}{\bm{#1}}}\xspace}

\DeclareMathOperator{\hAssumeOp}{assume}
\DeclareMathOperator{\hAssertOp}{assert}
\DeclareMathOperator{\hAssignOp}{assign}
\DeclareMathOperator{\hCallOp}{hw\_call}
\DeclareMathOperator{\hWithOp}{with}
\DeclareMathOperator{\hWhereOp}{where}
\DeclareMathOperator{\hByOp}{by}
\DeclareMathOperator{\hEnsuringOp}{invariant}
\DeclareMathOperator{\hIfOp}{if}
\DeclareMathOperator{\hElseOp}{else}
\DeclareMathOperator{\hForOp}{for}
\DeclareMathOperator{\hPforOp}{par-for}

\DeclareMathOperator{\hAparamsOp}{app-params}
\DeclareMathOperator{\hHparamsOp}{hw-params}
\DeclareMathOperator{\hDparamsOp}{deriving}
\DeclareMathOperator{\hForallOp}{\forall}
\DeclareMathOperator{\hLmapOp}{\forall}
\DeclareMathOperator{\hBufferOp}{Buffer}
\DeclareMathOperator{\hDcsOp}{layout-sketch}
\newcommand{\hAssume}{\asKword{\hAssumeOp}}
\newcommand{\hAssert}{\asKword{\hAssertOp}}
\newcommand{\hAssign}{\asKword{\hAssignOp}}
\newcommand{\hCall}{\asKword{\hCallOp}}
\newcommand{\hWith}{\asKword{\hWithOp}}
\newcommand{\hWhere}{\asLoopKword{\hWhereOp}}
\newcommand{\hBy}{\asLoopKword{\hByOp}}
\newcommand{\hEnsuring}{\asLoopKword{\hEnsuringOp}}
\newcommand{\hIf}{\asKword{\hIfOp}}
\newcommand{\hElse}{\asKword{\hElseOp}}
\newcommand{\hFor}{\asLoopKword{\hForOp}}
\newcommand{\hPfor}{\asLoopKword{\hPforOp}}

\newcommand{\hAparams}{\asKword{\hAparamsOp}}
\newcommand{\hHparams}{\asKword{\hHparamsOp}}
\newcommand{\hDparams}{\asKword{\hDparamsOp}}
\newcommand{\hForall}{\asDataKword{\hForallOp}}
\newcommand{\hLmap}{\asDataKword{\hLmapOp}}
\newcommand{\hBuffer}{\asDataKword{\hBufferOp}}
\newcommand{\hDcs}{\asDataKword{\hDcsOp}}

\newcommand{\hTaccL}{\asDataKword{\bm{[}}}
\newcommand{\hTaccR}{\asDataKword{\bm{]}}}

\newcommand{\hTacc}[2]{#1 \hTaccL #2 \hTaccR}

\newcommand{\hTaccSep}{\asDataKword{\bm{,}}}

\newcommand{\IE}{i.e.,\xspace}
\newcommand{\EG}{e.g.,\xspace}
\newcommand{\ETC}{etc.\xspace}

\AtBeginDocument{%
  }

\acmISBN{978-1-4503-XXXX-X/2018/06}

\begin{document}

\title{Verification of Compiler-to-Accelerator Mappings for Machine Learning Accelerators}


\author{Akash Gaonkar}
\affiliation{
    \institution{Princeton University}
    \city{Princeton}
    \state{NJ}
    \country{USA}
}
\email{agaonkar@princeton.edu}

\author{Mike He}
\affiliation{
    \institution{Princeton University}
    \city{Princeton}
    \state{NJ}
    \country{USA}
}
\email{mikehe@princeton.edu}

\author{Yi Li}
\affiliation{
    \institution{Meta Platforms}
    \city{Menlo Park}
    \state{CA}
    \country{USA}
}
\email{yili.leeoh@gmail.com}

\author{Bo-Yuan Huang}
\affiliation{
    \institution{Intel INT31}
    \city{Santa Clara}
    \state{CA}
    \country{USA}
}
\email{byhuang116@gmail.com}

\author{Andrew Cheung} 
\affiliation{
    \institution{University of Washington}
    \city{Seattle}
    \state{WA}
    \country{USA}
}
\email{acheung8@cs.washington.edu}

\author{Vishal Canumalla}
\affiliation{
    \institution{Stanford University}
    \city{Palo Alto}
    \state{CA}
    \country{USA}
}
\email{vjc@stanford.edu}

\author{Gus Henry Smith} 
\affiliation{
    \institution{University of Washington}
    \city{Seattle}
    \state{WA}
    \country{USA}
}
\email{guscomps@gmail.com}

\author{Zachary Tatlock}
\affiliation{
    \institution{University of Washington}
    \city{Seattle}
    \state{WA}
    \country{USA}
}
\email{ztatlock@cs.washington.edu}

\author{Grigory Fedyukovich}
\affiliation{
    \institution{Florida State University}
    \city{Tallahassee}
    \state{FL}
    \country{USA}
}
\email{grigory@cs.fsu.edu}

\author{Sharad Malik}
\affiliation{
    \institution{Princeton University}
    \city{Princeton}
    \state{NJ}
    \country{USA}
}
\email{sharad@princeton.edu}

\author{Aarti Gupta}
\affiliation{
    \institution{Princeton University}
    \city{Princeton}
    \state{NJ}
    \country{USA}
}
\email{aartig@cs.princeton.edu}





\begin{abstract}
To meet the performance needs of modern machine learning (ML) applications, ML compiler frameworks
support \emph{\mappings} that offload parts of application code to operations in specialized hardware accelerators.
However, most of these frameworks do not verify
these mappings down to the hardware level, potentially resulting in functional mismatches.
In this paper we propose \sys, the first framework for formally verifying the correctness of \mappings for coarse-grained intrinsics in ML accelerators, with respect to a formal hardware semantics. 
\sys does not require additional information from the compiler, and verifies the functional equivalence of the application code and the code for the mapped hardware accelerator intrinsic, including handling of complex loop nests and tensor data layouts in hardware. It effectively utilizes a pattern of 
\emph{aligning} software loops with the hardware, followed by \emph{relating} corresponding data layouts, to enable verification using well-aligned product programs. 
To support these steps, we propose two custom templates --- \emph{\skel, \dcs} --- to guide users in aligning loops and specifying data layout relationships, respectively.
We have developed a proof-of-concept prototype for \sys
and use it to successfully verify the correctness of several complex mappings for two recent open-source ML accelerators.
\end{abstract}





\maketitle

\begin{ignore}
\textbf{TODOs overall}
\begin{itemize}
    \item Add a descriptive legend to all figures.
    \item Capture "who" puts in the human effort and how much in figures, tables, text.
    \item Add Dafny and Why3 in related work discussions where needed (Intro, Hex section, Related work, etc.)
    \item \S1.3-4 Explicitly highlight when a thing is novel and/or when a thing is at least partially automated, in each of the steps. Want to track this as we write.
    \item \S7, report \#of check-sat calls rather than number of VCs, in \S7.2, report statistics measuring effort for VC-gen
    \item \S7.3 add column for QI
\end{itemize}

\textbf{TODOs @Grigory}
\begin{itemize}
  \item Implement invariant inference
  \item Writing \S6 (6.2 has new parts) -- see old text and supplementary/verify.tex
  \item Writing 1.3 -- from your perspective, identify the novelty/strength/insights in this work. What is the best way to put this in context of existing translation validation work you're familiar with.
\end{itemize}
\end{ignore}

\section{Introduction}
\label{s:introduction}

\begin{figure}
  \includegraphics[width=.85\columnwidth]{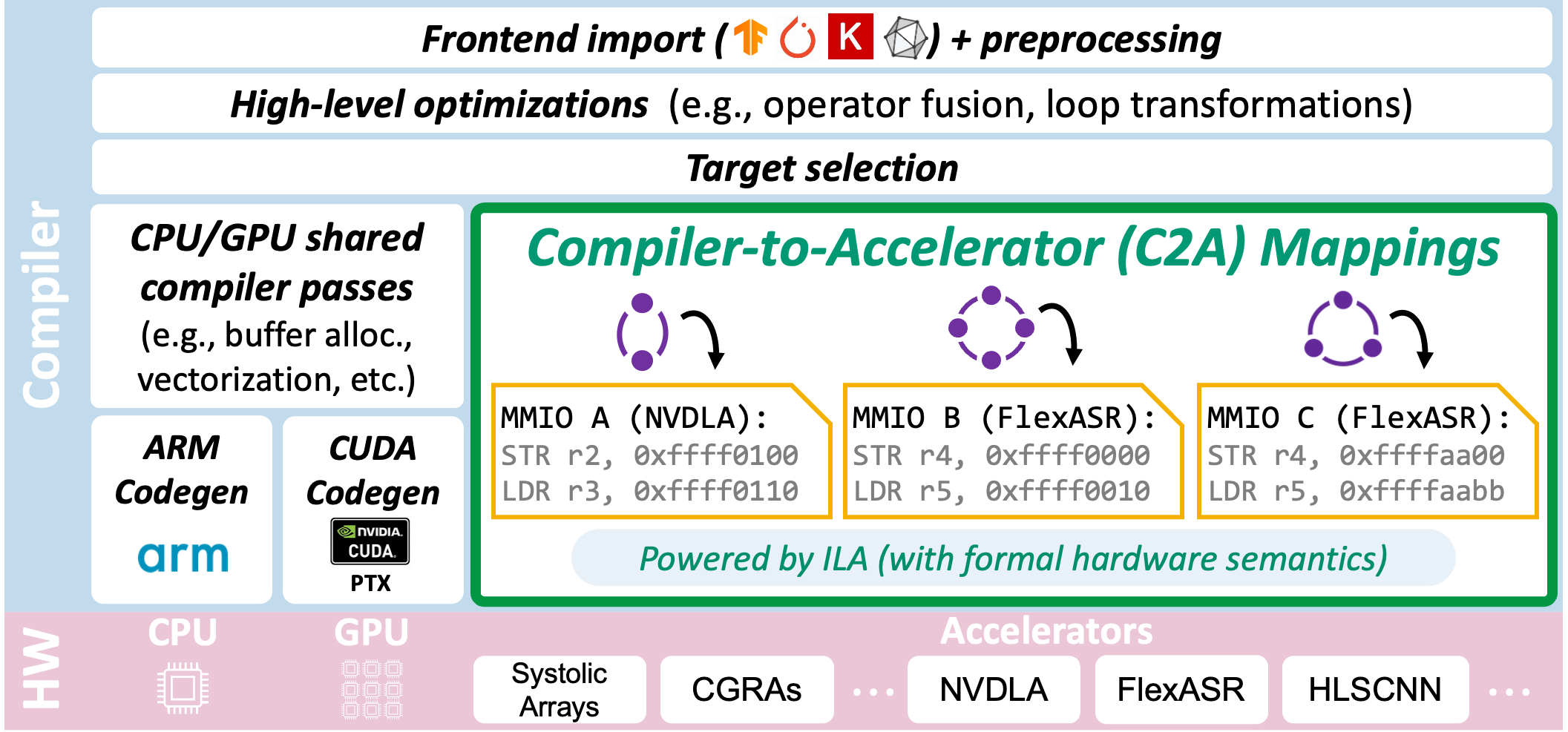}
  \vspace{-3mm}
  \caption{\textbf{A typical ML compiler flow that targets  hardware accelerators.} After high-level optimizations and target selection, code generation for hardware accelerators is performed by using \textbf{compiler-to-accelerator mappings} (green box).
  \sys verifies these mappings with respect to a formal hardware semantics.
  }
  \label{fig:c2a}
  \vspace{-3mm}
\end{figure}

\hideableoutline{
\begin{itemize}
  \item To support the demands of modern ML applications, recent compiler frameworks support compiling application code to specialized hardware accelerators. To target the coarse-grained intrinsics provided by these accelerators, frameworks use \emph{\mappings (\camappings)}.
  \item Application developers using a compiler framework expect compiler correctness, \IE that the compiler output is \emph{semantically equivalent} to the original application program. While existing frameworks verify a variety of code transformations, \camappings have not been verified in prior work. \akashc{except a case study in Pono and for a family of accelerators known as \emph{coarse-grained reconfigurable arrays (CGRAs)}.}
  \item We propose \sys (\underline{B}ridging \underline{O}paque \underline{L}oops and \underline{T}ensors), a novel framework and the \emph{first} one (to the best of our knowledge)
for formally verifying the correctness of \mappings from ML application code to coarse-grained accelerator intrinsics.
  While these mappings are a small part of a larger compiler flow (Fig.~\ref{fig:c2a}), reasoning about them is necessary to ensure compiler correctness with respect to a formal hardware semantics.
\end{itemize}}

Modern machine learning (ML) applications continually demand better performance (speed, power usage, etc.) that is being met through specialized accelerators. 
To support this computation trend, recent ML compiler frameworks such as BYOC~\cite{chen2021byoc}, Exo~\cite{ikarashi2022exo}, MLIR~\cite{lattner2021mlir}, Halide~\cite{ragankelley2013halide,liu2023stencils}, and 3LA~\cite{huang2024_3la} 
can compile application code to run on these specialized hardware accelerators in addition to the CPU and GPU.
%
Hardware accelerators provide \emph{intrinsics} for specialized coarse-grained operations, such as matrix multiplication~\cite{genc2021gemmini, moreau2019hardwaresoftware}, or linear-layer~\cite{tambe2021flexasr}.  This specialization is made possible by significant optimizations in the hardware accelerator implementations, particularly by: (i) using hardware parallelization techniques to expedite loops, and (ii) customizing where and how tensor data are stored in accelerator memory.
Accordingly, many frameworks use \emph{\mappings} for coarse-grained intrinsics, to produce highly performant executables and/or runtimes.

This context is illustrated in Fig.~\ref{fig:c2a}.
An ML compiler (such as Exo~\cite{ikarashi2022exo}, Mosaic~\cite{bansal2023mosaic}, or 3LA~\cite{huang2024_3la}) uses various techniques for code optimization and target selection on the compiler IR (intermediate representation). 
Then, during code generation, it maps pieces of IR code 
to some hardware accelerator intrinsics using a \mapping (abbreviated \camapping, or simply a mapping) -- three such mappings are shown pictorially in the figure.
Intrinsics are invoked by the host processor through MMIO (Memory-Mapped IO) or similar instructions -- we refer to these as MMIO code in the rest of the paper. Executing the MMIO code by the host processor is essentially like calling a function that is implemented in hardware. 

\subsection{Motivation: Verifying correctness of \mappings}

Aside from performance, the \emph{correctness} of the compiled code is the compiler’s primary requirement.
That is, the compiler's output must be \emph{equivalent} to the original program.
Many existing ML compilers~\cite{liu2022atl,bansal2023mosaic,courant2021polyverif,ikarashi2022exo,clement2022halideverif,melchert2025cgraverif} use \emph{and verify} code transformations such as loop reordering, loop tiling, \ETC done in software (discussed in detail \S\ref{s:related-work}). However, most stop short of considering a formal hardware semantics, and the step of mapping a code fragment to a coarse-grained hardware accelerator intrinsic (i.e., the MMIO code for this hardware function call) is often not verified. 
If there is a mistake in configuring the accelerator or there is a functional mismatch with the supported intrinsic, then errors 
can result from a single invocation of the accelerator intrinsic (and accumulate over multiple invocations in an end-to-end ML application).  

\myparagraph{Problem definition.} Given a \mapping from some \cspf to some \aspfue,
our goal is \emph{to prove the functional 
equivalence of the code on the two sides, including a formal hardware semantics}, \IE given the same input tensors, they should produce the same output tensors. We consider the general case of a \emph{black-box mapping}, \IE where there may not be any internal compiler steps with additional information about how the mapping was constructed or selected. In this general form, this can be viewed as the classic program equivalence problem. This general case is also useful for a third-party independent verification of a mapping.

The problem of verifying the correctness of a \mapping has some similarities with translation validation~\cite{pnueli1998translationvalidation,necula2000translationvalidation}, where programs before and after some compilation step are checked for equivalence. However, the programs that we aim to verify for equivalence constitute the mapping itself, which is used by an ML compiler during code generation. Thus, by verifying the equivalence of these programs, we prove correctness of the associated compiler step.

\akash{
\ag{We would like to emphasize that the goal of our work}  is to demonstrate the \emph{tractability} of program-equivalence based verification in this context, rather than \emph{full automation} (which would require more engineering effort in developing a tool chain that supports an existing compiler).
The reader may wonder why \mappings are challenging to verify correct---many hardware intrinsics have been verified through standard hardware verification techniques, such as bounded model checking (BMC)~\cite{bmc} ---so why not the \aspfue of the mappings shown in Fig.~\ref{fig:c2a}?
}
\ag{The main reason is that mappings to coarse-grained accelerators are hard to verify because the hardware operations contain \emph{many and deeply nested loops} (due to hardware parallelism) and \emph{complex data layouts} (in hardware memory buffers). Techniques based on BMC are not able to fully unroll the loops within resource limits, and existing invariant-based techniques (using interpolants~\cite{zhang2018ilamcm} or IC3~\cite{bradley2011ic3}) do not scale on large complex buffers.}
\ag{In \S\ref{s:intro-challenges}, we describe the verification challenges due to these hardware features in detail, }
\akash{and reveal them in a real-world example in \S\ref{s:overview-example}. We will demonstrate how we have customized program equivalence techniques 
in this hardware-driven context. 
We successfully verify several \camappings, even as standard techniques do not scale (\S\ref{s:evaluation-experiments}).}

\subsection{Proposed framework: \sys}
\label{s:intro-framework}

In this paper, we propose \sys
(\underline{B}ridging \underline{O}paque \underline{L}oops and \underline{T}ensors), a 
new framework and the \emph{first} one for formally verifying the correctness of \mappings for coarse-grained ML accelerator intrinsics, \emph{with respect to a formal hardware semantics}. 
\sys requires the semantics of the IR code and the MMIO code, but does not depend upon any compiler information such as internal correspondences between smaller fragments in the IR code and sub-operations or internal buffers in the hardware.

There are other efforts that have taken a different approach for considering hardware semantics when compiling ML applications to hardware accelerators. In particular, a recent effort~\cite{melchert2025cgraverif} verifies an ML compiler pipeline down to hardware accelerators.
It is based on a custom compiler that maps software to a restricted family of accelerators (called coarse-grained reconfigurable arrays) in a series of smaller compilation steps. They use translation validation on these smaller steps, using information from the compiler (e.g., scheduling, memory mapping) to perform verification.
Another recent approach~\cite{pouchet2024hlsverif} uses High-Level Synthesis (HLS), where source-to-source transformations on C/C++ programs are used for optimizing ML operations (such as matrix-multiply). Here, the hardware generated by the HLS compiler (\EG a systolic array accelerator) from the optimized program is (presumably) correct-by-construction. 
While these synthesis-based approaches greatly help verification of the ML compilation pipeline, their goals and approaches are incomparable to ours -- their compilers do not target existing/custom hardware accelerator designs (such as FlexASR~\cite{tambe2021flexasr}) or coarse-grained ML operations beyond those that their compilers can automatically synthesize to reconfigurable/systolic arrays. 

\subsection{Verification Challenges}
\label{s:intro-challenges}
There are many challenges in verifying the correctness of \mappings. An ML compiler typically uses a tensor program representation as its high-level IR,
and explores different loop reorderings, loop tilings, \ETC, to optimize for the most performant loop structure, commonly referred to as the \emph{loop nest}. During code generation, \camappings encapsulate the mapping from selected IR code to MMIO code for the accelerator. 

\myparagraph{Formal hardware semantics for MMIO code.} A core verification challenge is to bridge the gap between software semantics on the IR side with hardware semantics (typically state transition systems) on the other. As mentioned earlier, many existing ML compiler efforts~\cite{liu2022atl,ikarashi2022exo,lattner2021mlir,chen2021byoc} stop short of verifying against a formal hardware semantics. Our first challenge is that we need to capture the hardware semantics of the MMIO code in a given mapping.

\myparagraph{Complex loop nests and tensor data layouts in hardware.} 
In coarse-grained \camappings, the hardware side often has complex loop nests due to parallel processing elements and/or single-instruction-multiple-data (SIMD) parallelism, in addition to loops using hardware counters. Furthermore, the correspondences between tensor data in the IR code and the accelerator code are obscured by highly optimized \emph{tensor data layouts} within an accelerator memory. In particular, partitioning, rearrangement, and duplication of tensor data in hardware introduces complicated address arithmetic. Thus, checking equivalence of corresponding outputs requires relating complex data layouts on the two sides, but users often struggle to define them and verifiers struggle to reason about them. 

\subsection{\sys: Key Ideas and Contributions}
\label{s:intro-contributions}

\akash{\textbf{Insight \#1:} Our first insight is that mappings to coarse-grained accelerators are hard to verify because the hardware operations contain \emph{many and deeply nested loops} (due to hardware parallelism) and \emph{complex data layouts} (in hardware memory buffers). Frustratingly, these loop nests and data layouts are opaque within the semantics of the \aspfue.}

To address the core challenge of capturing the formal hardware semantics and bridging the gap with IR code, we have developed an \emph{Intermediate Verification Language (IVL)} called \lang. We do not claim novelty in its design, which is inspired by other IVLs such as Dafny~\cite{leino2010dafny} for program verification, and HLS languages (such as synthesizable C/C++, SystemC~\cite{panda2001systemc}) that model hardware. 

We use \lang to represent the MMIO code as a stateful program (detailed in \S\ref{s:bolt-hex}), where the formal hardware semantics is specified by an Instruction Level Abstraction (ILA) model~\cite{huang2018ila}.
Essentially, an ILA model specifies a  state transition update for each operation (MMIO command) of the accelerator, where the update is hierarchically defined by a sequence of updates to some software-visible architectural state. Importantly, these state transition updates specify updates due to hardware-implemented parallel as well as iterative loops. 
In addition to explicating hardware semantics, \lang provides features for customized modeling of loop nests and tensor buffers touched in the \aspfue, since these play a critical role in verification. Also, it provides support for users to specify annotations that enable automated verification via Satisfiability Modulo Theories (SMT) solvers (e.g., Z3~\cite{demoura2008z3}).

For verifying program equivalence, \sys leverages known techniques~\cite{barthe2011relational,zaks2008crossproduct,sousa2016cartesian}, where a \emph{``product''} program is constructed over the two given programs to check that they produce the same outputs when given the same inputs.
There has been extensive work on \emph{aligning} corresponding parts (\EG loops) of the given programs, to enable use of \emph{relational invariants} in the product program for decomposing and simplifying  verification~\cite{terauchi05secure,barthe2011relational,zaks2008crossproduct,sousa2016cartesian,angelis2016relationalverif, unno2021constraintbased,GrigoryLpar17,DBLP:conf/tacas/HamzaF23}.
However, none of the existing techniques (detailed in \S\ref{s:related-work}) have been applied to \camappings or to loop nests of similar complexity as in ML accelerator mappings.  


Inspired by prior work, the key idea in \sys is to first \emph{align} the loop nests on the software side with the hardware side, and then \emph{relate} data layouts in the tensor updates in the corresponding loops. In our setting, we have found this pattern --- \emph{align-loops-with-hardware} and then \emph{relate-layouts-of-buffers} --- to be very useful. It enables decomposing the overall verification problem, which is essential in practice for verification to succeed. 
\akash{We have developed some insights supporting these two steps, which we have incorporated into the key components in the \sys framework, depicted in Fig.~\ref{fig:methodology} (and detailed in the next section with a motivating example).}

\akash{\textbf{Insight \#2 (Align step)}: To support align-loops-with-hardware, we only need to align software with the \emph{compute-performing loops in the hardware side.} 
First, it is easier to perform rewriting on the pure-functional IR code than on the stateful accelerator Hex code.
Furthermore, data movement loops in hardware do not affect the \emph{functional} behavior of the hardware side and can be handled in a later step (Relate). These insights lead us to introduce a template called a \emph{\skel}, which extracts tensor computing loops in the hardware-based operation (including those due to parallelization or counter-based iteration). The \skel is used as a guide for the user to rewrite the IR code to align its loops with hardware. from the latter, and reduce the number of loops that should be aligned (ideally) in the former.}

\akash{\textbf{Insight \#3 (Relate step):} For relate-layouts-of-buffers, we combine information (shared loop counters, tensor sizes, buffer overwriting) provided by the software IR and hardware sides. Concretely, we introduce a template called a \emph{\dcs}, which identifies a correspondence between the tensor outputs, assuming correspondences on the tensor inputs. This provides a flexible template for each loop, capturing iterative updates in data computation and data movement in the loop, which is used to construct important relational invariants that help verification succeed. While the development of these templates are user-driven, we provide \lang-based 
tactics to guide the user. These steps enable construction of a \emph{well-aligned product program}, along with annotations for data layout invariants and other invariants, 
which is verified using automated SMT-based techniques.}


\begin{figure}[t]
  \includegraphics[width=0.88\columnwidth]{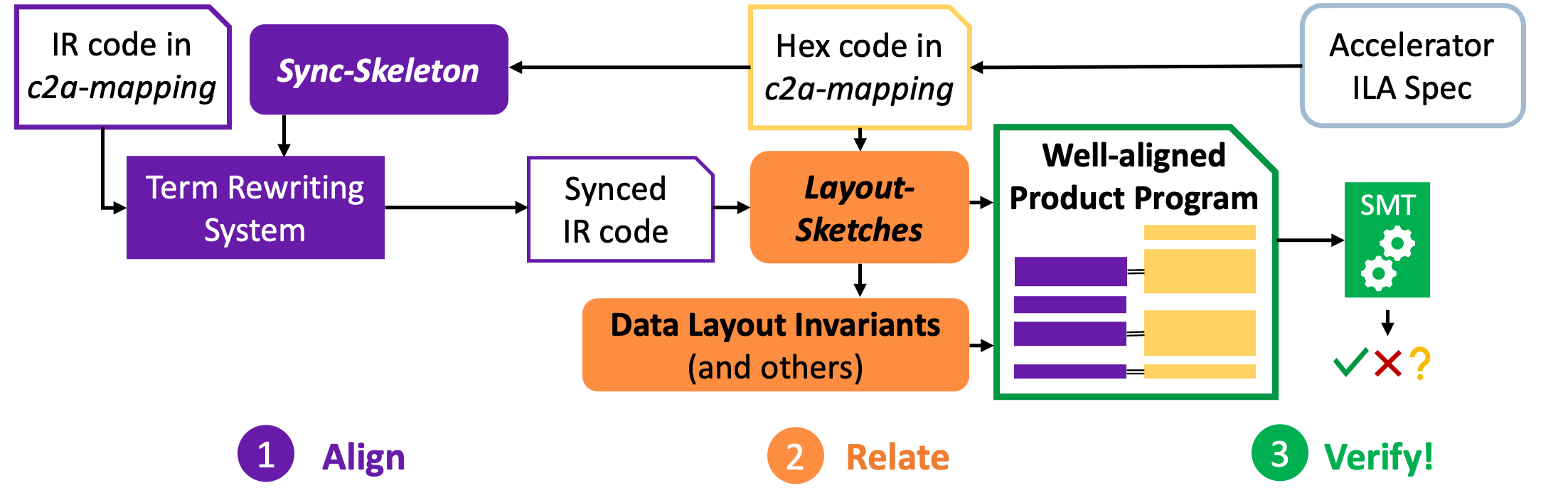}
  \vspace*{-2mm}
  \caption{\sys Framework: Components and Main Steps.}
  \label{fig:methodology}
  \vspace*{-1mm}
\end{figure}

\myparagraph{\sys Prototype.} We have developed a proof-of-concept prototype, where the user provides the \cspf and the \aspf, interacts with the prototype to develop the custom templates \skel and \dcs, performs IR rewriting guided by the \skel, and provides loop annotations based on the \dcs. 
The verification procedure in \sys is automated and uses the Z3 SMT solver~\cite{demoura2008z3}.
\akash{We log the user effort involved in our prototype as part of our evaluation (\S\ref{s:evaluation-key-stats}). At each step, the user largely follows the procedures we prescribe, setting the stage for future automation.}


\pagebreak
\myparagraph{Summary of Contributions}
\begin{itemize}
\item We present \sys, the \emph{first} framework for verifying correctness of \mappings for coarse-grained intrinsics in compilers for ML accelerators, with respect to a formal hardware semantics. We consider the general case where no information is available from the compiler about how these mappings are constructed, and prove program equivalence between the software and hardware code in the mapping. 

\item We leverage techniques based on product programs and propose custom templates for guiding a user to align loops and specify data layout relationships --- the \skel and \dcs, respectively. We provide heuristics and tactics for extracting these templates from programs written in \lang, an intermediate verification language that powers \sys, and for constructing relational invariants that help the solver's performance. \akash{In the common case, these heuristics and tactics reduce the burden of aligning loops and specifying relational invariants to step-by-step procedures.}

\item 
We evaluate \sys on case studies from two recent open-source ML accelerators~\cite{tambe2021flexasr,whatmough2019hlscnn} with available formal hardware semantics. These case studies show \sys's capabilities in successful verification of several complex \mappings. 
Notably, our most complex mapping --- linear-layer for the FlexASR accelerator --- has 29 loops in the product program (many nested, up to maximum depth 7). As far as we know, none of the existing program equivalence techniques/tools (based on either automated reasoning or interactive theorem proving) has demonstrated successful verification of such complex product programs.
\end{itemize}

\begin{ignore}
In this paper, we propose \sys
(\underline{B}ridging \underline{O}paque \underline{L}oops and \underline{T}ensors), a 
new framework and the \emph{first} one 
for formally verifying the correctness of \mappings from ML application code to coarse-grained hardware accelerator intrinsics. 

These mappings are essentially used for code generation, as shown in the green box in an illustrative compiler flow depicted in Fig.~\ref{fig:c2a}. Although they constitute a small part of a larger ML compilation flow (including high level optimizations, target selection, \akash{etc.}) 
reasoning about them is critical for ensuring correctness with respect to a formal hardware semantics.
We demonstrate that \sys 
can successfully verify the correctness of several complex mappings for two state-of-the-art ML accelerators~\cite{tambe2021flexasr,whatmough2019hlscnn}. (We plan to make our mappings and programs publicly available.)
\end{ignore}

\begin{ignore}
    
\subsection{Verification Problem and Challenges}

\hideableoutline{
\begin{itemize}
  \item Terminology: tensor program, loop nest, \camappings = code generation mappings from IR code to accelerator intrinsics, \cspf, MMIO code, and later on, \aspf.
  \item \emph{Our goal is to prove semantic equivalence between the \cspf and the \aspf}, \IE given the same inputs, both programs will generate the same outputs based on their respective tensor/hardware semantics.
  \item As a running example, Fig~\ref{fig:intro-mapping} depicts a \camapping for an accelerator called FlexASR~\cite{tambe2021flexasr}. It depicts the \cspf, MMIO code \akash{(TODO)}, and \aspf for performing a linear-layer operation using FlexASR.
  \item Note the complex loop nests and data layout within the hardware, as depicted in Fig.~\ref{fig:intro-mapping}.
\end{itemize}}

An ML compiler typically uses a \emph{tensor program} representation as its high-level Intermediate Representation (IR), and explores different loop reorderings, loop tilings, \ETC, to optimize for the most performant loop structure, commonly referred to as the \emph{loop nest}.
%
%
%
During code generation, \camappings encapsulate the mapping from selected IR code to MMIO code for the accelerator.

\myparagraph{Problem definition.} Given a \camapping from some \cspf to some \aspfue,
\emph{our goal is to prove the functional 
equivalence of the code on the two sides}, \IE given the same input tensors, they should produce the same output tensors. In its general form, where we do not depend on additional information about the mapping, this can be viewed as the classic program equivalence problem.

The problem of checking a \camapping has some similarities with translation validation~\cite{pnueli1998translationvalidation,necula2000translationvalidation} approaches, where programs before and after some compilation step are checked for equivalence. However, note that the programs that we check for equivalence constitute the mapping itself, which is used by a compiler during code generation. Thus, by verifying equivalence of the programs in a given mapping, we prove correctness of the associated compiler step. 

\begin{figure}
  \noindent\makebox[\textwidth][c]{%
  \input{figures/flin-mapping-huge}}
  \vspace*{-7mm}
  \caption{Running example: the \mapping for a linear-layer operation in FlexASR.}
  \vspace*{-3mm}
  \label{fig:intro-mapping}
\end{figure}

To verify program equivalence, we aim to leverage existing techniques based on product programs~\cite{barthe2011relational,zaks2008crossproduct,angelis2016relationalverif}. 
However, there are many challenges in our setting. 
To illustrate these challenges concretely, we consider a running example of a \camapping shown in Fig.~\ref{fig:intro-mapping} for a recent open-source
ML accelerator called FlexASR~\cite{tambe2021flexasr}.
This mapping offloads a linear-layer operation to an intrinsic in FlexASR, as used in the 3LA compiler~\cite{huang2024_3la}. 

\myparagraph{\Cspf for linear-layer.} The purple box (upper left) shows the \cspf, in a basic pure-functional tensor language that uses \emph{maps} and \emph{reduces} to perform tensor computations.
Note that these are highly structured loop operations that maintain multidimensional tensor representations for data accesses. In particular, the notation $\csMap{i=0}{n} e$ constructs a tensor by mapping each $i \in \{0, \ldots, n-1\}$ to the result of $e$ at $i$, \IE $[e|_{i=0} \cdots e|_{i=n-1}]$, and the notation $\csRed{i=0}{n}(s = e_1 \csIn e_2)$ accumulates the value of $s$, starting at $s = e_1$ and updating $s = e_2|_{i, s}$ for each successive value of $i \in {0, \ldots, n-1}$.

\myparagraph{\Aspfue for FlexASR linear-layer.}
The yellow box on the bottom left in Fig.~\ref{fig:intro-mapping} shows the \aspfue for performing a linear-layer using the FlexASR accelerator, where the host processor is an Arm Cortex-A53 (supporting A32 assembly). 
Several MMIO commands configure the accelerator before the final command (emphasized in red) triggers the linear-layer computation within the accelerator hardware.

\myparagraph{Challenge: Formal hardware semantics?} The first challenge is that we need to capture the formal hardware semantics of the MMIO code, in order to check equivalence against the functional IR program. To address this challenge, we have developed an \emph{Intermediate Verification Language (IVL)} called \lang. We do not claim novelty in its design, which is inspired by other IVLs such as Dafny~\cite{leino2010dafny} for program verification, and HLS languages (such as synthesizable C/C++, SystemC~\cite{panda2001systemc}) that model hardware. We use \lang to represent the MMIO code as a stateful program (detailed in \S\ref{s:bolt-hex}), where the formal hardware semantics is specified by an Instruction Level Abstraction (ILA) model~\cite{huang2018ila}.
Essentially, an ILA model specifies a  state transition update for each operation (MMIO command) of the accelerator, where the update is hierarchically defined by a sequence of updates to some software-visible architectural state. Importantly, these state transition updates specify updates due to hardware-implemented parallel as well as iterative loops. 
In addition to explicating hardware semantics, \lang provides features for customized modeling of loop nests and tensor buffers touched in the \aspfue, since these play a critical role in verification. Furthermore, it  provides support for users to specify annotations that enable automated verification via Satisfiability Modulo Theory (SMT) solvers (e.g.,  Z3~\cite{demoura2008z3}). 

\myparagraph{\Aspf for FlexASR linear-layer.}
The yellow box on the right in Fig.~\ref{fig:intro-mapping} shows the accelerator code 
for the running example in \lang. 
Each \hCall statement here (lines 6-10) corresponds one-to-one with an MMIO command, 
after a preamble that defines \emph{parameters} (lines 1-5) that represent key dimensions and sizes for configuring the accelerator.
%
The code block in lines 10-27 shows (a part of) the computation triggered due to the operation 
in line 10.

Note that the FlexASR accelerator code has \emph{27 hardware loops} that implement the linear-layer operation! These are represented in \lang via imperative-style loops, where the \hPfor loops capture 
hardware parallelism due to parallel processing elements (\EG line 11) or single-instruction-multiple-data (SIMD) parallelism (\EG lines 16, 21).
Note also that \lang captures memory accesses and assignments (\EG lines 17, 23-25) that manipulate tensor data in hardware. FlexASR often uses significant address arithmetic (\EG line 17) to optimize tensor data layouts in its memory buffers.

\myparagraph{Verification Challenges:} A core verification challenge in any ML compiler flow is to bridge the gap between software semantics on one side (typically functional tensor programs) with hardware semantics (typically state transition systems, translated to imperative stateful programs) on the other. Many existing efforts~\cite{liu2022atl,ikarashi2022exo,lattner2021mlir,chen2021byoc} stop short of modeling the formal hardware semantics. 

In the specific context of coarse-grained \camappings, such as our running example, there is no information about a decomposition into smaller compilation steps,  
and we need to verify equivalence of two large complex programs with potentially many and deeply nested loops.
Furthermore, the correspondences between tensor data in the IR code and the accelerator code are obscured by highly optimized \emph{data layouts} within an accelerator memory. For instance, partitioning, rearrangement, and duplication of tensor data in hardware introduces complicated address arithmetic. Thus, checking equivalence of corresponding outputs requires relating complex data layouts on the two sides, 
but users often struggle to define them and SMT solvers struggle to reason about them. 

\end{ignore}

\begin{ignore}
    
\subsection{Verification of \camappings}

\akash{Verifying a \camapping is a} translation validation problem, where we focus on establishing the equivalence of the code before and after the \camapping is applied. However, there are two important differences from existing translation validation \akash{work} (discussed in \S\ref{s:related-work}). First, our mappings for ML applications bridge the gap between software semantics on one side (typically functional tensor programs) with hardware semantics (typically state transition systems, translated to imperative stateful programs) on the other.
To the best of our knowledge, existing \akash{translation validation} efforts in ML compilers \akash{cover much smaller semantic gaps and typically do not reason about hardware semantics.}
Second, the mapping is usually a \emph{black-box one-shot substitution} and could involve large complex \cspf and coarse-grained \aspf, \EG an entire linear layer (as shown in Fig.~\ref{fig:intro-mapping}).
Thus, there is no information about a decomposition into smaller compilation steps, across which the programs could be verified for equivalence. More critically, due to the complex loop nests and tensor data layouts in these programs, verifying their equivalence is very challenging for existing approaches, as highlighted below.

\myparagraph{(Challenge C1) Many nested loops.}
To support coarse-grained accelerator intrinsics,
a \mapping often captures a large tensor computation containing many nested loops.
%

\myparagraph{(Challenge C2) Loop alignment is difficult.}
One can consider applying program equivalence techniques based on loop alignment
after synchronizing the two sides into a \emph{``well-aligned''} product program~\cite{barthe2011relational,zaks2008crossproduct,sousa2016cartesian}.
However, loop nests in the accelerator fragment, especially those due to \emph{hardware} data-parallelism and tensor data movement make it difficult to find alignment.
%

\myparagraph{(Challenge C3) Complex data layouts.}
The correspondences between tensor data in the \cspf and \aspf are obscured by highly optimized \emph{data layouts} within an accelerator memory, where partitioning, rearrangement, and duplication of tensor data often introduces complicated address arithmetic. Relating complex layouts typically requires quantified invariants, which users struggle to define and SMT solvers struggle to reason about.
%

\end{ignore}

\begin{ignore} 

\subsection{\sys Framework: Overview and Key Ideas}

\hideableoutline{
\begin{itemize}
  \item We leverage ILAs as \emph{verifiable specifications of hardware} that capture a formal state transition semantics for the accelerator intrinsics.
  \item We leverage techniques based on product programs to check program equivalence, which we extend to handle coarse-grained accelerator intrinsics and loop nests of the complexity found in mappings for ML applications.
  \item The main components of \sys are shown in Fig~\ref{fig:methodology}. We introduce an intermediate verification language called \lang, which captures the loop nests and data layout within a hardware design for the purpose of reasoning about their equivalence to software loops and tensors. Our framework introduces distinct steps to \emph{Align} loop nests, \emph{Relate} tensor data layout, and \emph{Verify} equivalence using a well-aligned product program. Primitives of \lang enable the construction of two templates that guide aligning loops (the \skel during Align) and defining relational loop invariants about data layout (the \dcs during Relate). \akashc{In text, we will describe each step individually.}
\end{itemize}}

\begin{ignore}

\textbf{FOR REFERENCE}

- Hex language features: 
(a) hardware-based loop nests 
-- hardware data parallelism: PE, SIMD
-- iteration using a counter
Both could be used for computation or data movement.
(b) multidimensional buffers
-- specification of a layout sketch, which expresses relationships between address arithmetic
(partitioning, duplication, and rearrangement of tensor data introduces complex address arithmetic)

- Sync-skeleton
We provide a heuristics-based procedure to extract a sync-skeleton from an \aspf, where each loop in the sync-skeleton identifies a tensor (map/reduce) computation, and abstracts away loops with only data movement. The sync-skeleton guides a user to perform rewriting on the \cspf to align the loop nests on both sides, in the Align step of Bolt. 

- Rewriting rules 
We provide a system of rewriting rules, standard and customized based on the HEX code (e.g., parameter substitution rules), for a user to align the loops in the \cspf to those in the sync-skeleton. The rewriting and alignment is best-effort, and any un-aligned loops on either side may require additional invariants but do not affect the soundness of verification.

- Data layout sketches
We provide tactics for a user to specify a data layout sketch in HEX for each aligned loop. A data layout sketch identifies the correspondence between the tensor elements in the \cspf and the buffer elements in the \aspf that are updated due to assignments in the loop, respectively, assuming some given correspondence between the inputs to the assignment. In the Relate step of Bolt, we provide a procedure to construct relational invariants for each aligned loop from the data layout sketches. 

- Verification with quantified invariants
The Align and Relate steps in Bolt enable construction of a well-aligned product program. We then use existing program verification techniques to generate Verification Conditions (VCs) based on relational invariants at aligned loop boundaries,  
which are verified using an off-the-shelf SMT solver such as Z3~\cite{demoura2008z3}.
We also provide an automated quantifier instantiation procedure to help a solver verify the \camapping given quantified invariants constructed from data layout sketches. 

\textbf{END REFERENCE}
\end{ignore}

%
\sys is a framework designed to support verification of  given \camappings, by verifying equivalence between the \cspf and the accelerator code.
%
To check program equivalence, it leverages known techniques~\cite{barthe2011relational,zaks2008crossproduct,sousa2016cartesian}, where a \emph{``product''} program is constructed over the two given programs to check that they produce the same outputs when given the same inputs.
There has been extensive work on \emph{aligning} corresponding parts (\EG loops) of the given programs, to enable use of \emph{relational invariants} in the product program for decomposing and simplifying  verification~\cite{terauchi05secure,barthe2011relational,zaks2008crossproduct,sousa2016cartesian,angelis2016relationalverif, unno2021constraintbased,GrigoryLpar17}.
However, none of the existing techniques (detailed in \S\ref{s:related-work}) have been applied on \camappings or on 
loop nests of similar complexity as in ML accelerator mappings. 

\myparagraph{Key ideas in \sys.} Building on this approach, the key idea in \sys is to first \emph{align} the loop nests on the software side with the hardware side, and then \emph{relate} data layouts in the tensor updates in those loops. In our hardware setting, we have found this pattern of \emph{align-loops-with-hardware and relate-layouts-of-buffers} to be very useful in decomposing the overall verification problem, which is essential in practice for verification to succeed. To support the step of align-loops-with-hardware, we propose a template called a \emph{\skel}, which extracts tensor computing loops in the hardware-based operation (including those due to parallelization or counter-based iteration). For relating the layouts of tensor data in the buffers, we propose a template called a \emph{\dcs}, which identifies a correspondence between the tensor outputs, assuming correspondences on the tensor inputs. While the development of these templates is user-driven, we provide \lang-based heuristics and tactics to guide the user.

\begin{figure}[t]
  \includegraphics[width=0.88\columnwidth]{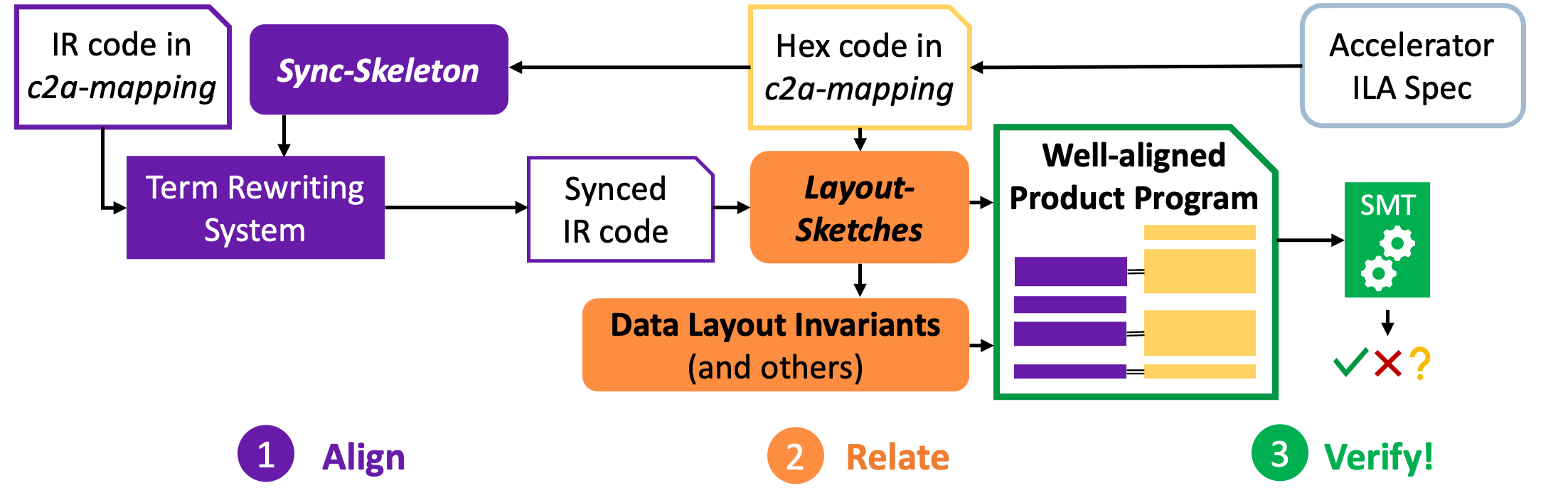}
  \vspace*{-2mm}
  \caption{\sys Framework: Components and Main Steps.}
  \label{fig:methodology}
  \vspace*{-1mm}
\end{figure}

An overview of \akash{\sys} 
is shown in Fig.~\ref{fig:methodology} and the main steps and components are briefly described below; a step-by-step illustration on the running example is 
in the next section (\S\ref{s:overview}).

\myparagraph{\lang language (\S\ref{s:bolt-hex}).} \sys is powered by the \lang IVL, which provides a common bridge between functional IR code on the one hand with state-based accelerator code that respects the formal hardware semantics (using the ILAs~\cite{huang2018ila}) on the other.
We view \lang as a repurposing of existing HLS-style languages (e.g., C/C++/SystemC), with custom features to support verification of hardware loop nests and complex data layouts.
In particular, \lang supports the development of \skel and \dcs (itself a first-class object),  
by identifying hardware counters and parameters as first-class objects, for reasoning about loop bounds and address arithmetic for data layouts.
Additionally, \lang is used to represent the product program and supports user annotations of relational invariants that help improve solver performance.  

\myparagraph{Loop alignment via the \Skel (\S\ref{s:bolt-align}).} 
The first main step (\emph{Align}) in \sys is to align the loops on the software side with those on the hardware side. 
We propose to guide this alignment via the \skel, a template that is extracted from the \aspf. We provide useful \lang-based  heuristics for extraction, such that each loop in the \skel identifies a tensor computation (a map/reduce operation) in the hardware while loops with only data movement are abstracted away. 
The reason for abstracting away the data movement loops is because they do not appear in the functional \cspf, so there is no need to try to align them. Note that these data movement loops \emph{will} be included at a later stage when the product program is constructed, since they determine tensor data layouts in the hardware buffers.

\myparagraph{Rewriting the \cspf for loop alignment (\S\ref{s:bolt-align}).}
The \skel guides a user to perform rewriting on the \cspf to align with the loop nests in hardware. (Since it is easier to rewrite functional programs, we advocate rewriting the \cspf, rather than rewriting the \aspf.) 
We provide a system of rewriting rules, standard and customized (based on \lang features, e.g., parameter substitution rules). A user can employ these rewriting rules to align the loops in the \cspf to those in the \skel. The rewriting and alignment is best-effort, and any un-aligned loops on either side may require additional invariants but do not affect the soundness of verification.

\myparagraph{Relating tensor correspondences via the \Dcs (\S\ref{s:bolt-relate}).}
After the loops have been aligned, the second main step (\emph{Relate}) in \sys is to relate the tensor data on the software side and the hardware side. 
This is done via a \dcs, a template that 
a user can specify in \lang for each aligned loop. Informally, a \dcs identifies the correspondence between the tensor elements in the \cspf and the buffer elements in the \aspf, respectively, that are updated due to assignments in the aligned loop, assuming some given correspondence between the inputs to the assignment. We provide useful tactics for users to create a \dcs, and also provide a simple procedure to construct relational invariants from a \dcs, to be added as annotations in the product program. 

\myparagraph{Verification via Product Programs (\S\ref{s:bolt-verify}).}
The Align and Relate steps enable construction of a \emph{well-aligned} product program in \lang, along with user annotations. In the third step (\emph{Verify}) in \sys, we use standard program verification techniques (e.g., similar to Dafny~\cite{leino2010dafny}) to automatically generate \emph{Verification Conditions (VCs)} based on relational invariants at aligned loop boundaries (and other invariants that a user may provide). These VCs are verified using an off-the-shelf SMT solver (Z3~\cite{demoura2008z3}).

\myparagraph{\sys Prototype.} We have developed a proof-of-concept prototype for the \sys framework, where the user provides the \cspf and the \aspf, interacts with the prototype to develop the custom templates \skel and \dcs in \lang, and provides annotations at aligned loop boundaries. The verification procedure in \sys is automated, and some simple invariants (details in \S\ref{s:bolt-relate}) are also automatically generated.

\begin{ignore}

We briefly highlight the key components of \sys and how they address the various challenges. (A step-by-step overview on the running example is described in the next section, \S\ref{s:overview}.)



\subsubsection{Leveraging product program verification.}
At a high-level, we leverage techniques based on product programs~\cite{barthe2011relational} to check program equivalence.
Essentially, these techniques construct a \emph{``product''} over the given programs, and then prove that it produces the same outputs when given the same inputs on both sides.
There has been extensive work on techniques that \emph{align} corresponding parts (\EG loops) of given programs, so that one can leverage \emph{relational invariants} in the product program to decompose and simplify verification~\cite{terauchi05secure,barthe2011relational,zaks2008crossproduct,sousa2016cartesian,angelis2016relationalverif, unno2021constraintbased,GrigoryLpar17}.
However, very few of these techniques (detailed in \S\ref{s:related-work}) handle coarse-grained accelerator intrinsics and loop nests of the complexity found in mappings for ML applications.

\end{ignore}

\begin{ignore}
Hex loops have annotations to capture SIMD-parallelism and hardware counters. 
These annotations help reason about hardware control logic, particularly with regard to address arithmetic,
as part of two heuristics we provide to users for extracting the sync-skeleton from the Hex code. The heuristics are conservative: they extract every loop that can correspond to a tensor map/reduce computation in the hardware, but may include extra loops that only perform data movement or be unsure if an extracted loop is a map/reduce. For our case studies, the sync-skeletons extracted using these heuristics successfully omitted all data-movement loops, and precisely identified most tensor map operations.

- Hex has customized support for quantified data layout invariants that are invertible, i.e., given the buffer address accessed, we can recover the values of the quantifiers. This is the typical case for tensors stored in memory buffers, as each address can only hold a single tensor element. We provide a novel quantifier instantiation procedure to automatically instantiate invariants of this type. When the data layout invariant was produced by a given layout-sketch, we further instantiate all data layout invariants of that layout sketch, as the input and output correspondences of the sketch share the same quantified range.

- Hex programs specify not only the key parameters used by the application and hardware but also the relationships between those parameters. These relationships inform parameter substitution rules within the Align-step term rewriting system, helping users to focus loop transformations to use hardware bounds and sizes, and therefore better align with the loops of the sync-skeleton. For our case studies, the rewrite rules we provide/extract enable a user to successfully align most loops of the sync-skeleton, and align all sync-skeleton loops when provided additional user-guidance on hardware numerical operations via custom-numeric rewrites.

\end{ignore}

\begin{ignore}

\subsubsection{Key ideas in \sys.}
An overview of the \sys framework and its main components is shown in Fig.~\ref{fig:methodology}.
\akash{As input we have a \camapping we wish to verify, consisting of given \cspf and \aspf.}
We propose the following steps in \sys to overcome the challenges C1-C3 described earlier.

\myparagraph{\stepT step.}
To address challenges C1 and C2, we propose the \emph{\defskeleton} (written \emph{\skel} in short), \akash{a template} extracted from the \aspf for a user-guided best-effort alignment of loops on the two sides.
\akash{In particular, we design a term rewriting system that enables users to} apply standard and customized program transformations on the \cspf to match the \skel.
Any unsynchronized loops may affect the performance, but do not affect the soundness of verification.

\myparagraph{\stepR step (\S\ref{s:bolt-relate}).}
To address challenge C3, we propose the \emph{\dcsfull} (written \emph{\dcs} in short), a template to help \akash{users} define quantified relational invariants that express dynamic relationships between data layouts in the \cspf and the \aspf at synchronized loop boundaries.

\myparagraph{\stepV step (\S\ref{s:bolt-verify}).} The above two steps enable us to construct a well-aligned product program in this step. We then use existing program verification techniques to
generate Verification Conditions (VCs) based on relational invariants at synchronized loop boundaries. \akash{We address solver limitations due to challenge C3 through a novel quantifier instantiation procedure, enabling the verification of VCs involving complex data layout using an off-the-shelf} SMT solver such as Z3~\cite{demoura2008z3}.
\\

\akash{Our contributions build on \lang, the intermediate verification language (IVL) we introduce to explicate and reason about loop nests and data layout within the hardware. Other IVLs such as Dafny~\cite{leino2010dafny} or hardware description languages such as SystemC could power the steps of \sys, but need customization respectively for capturing hardware loops or data layout invariants.}

Thanks to key ideas in \sys, our evaluations show successful verification of many complex mappings in two state-of-the-art ML accelerators. For example, the linear-layer mapping has 28 loops (many nested, up to maximum depth 7).

\end{ignore}

\subsection{Summary of Contributions}

\begin{itemize}
\item We propose \sys, the \emph{first} framework for verifying correctness of \mappings for coarse-grained intrinsics in compilers for ML accelerators, 
with respect to formal hardware semantics. We consider the general case where no information is available from the compiler about how these mappings are constructed. 
\item We leverage verification techniques based on product programs and propose custom templates for guiding a user to align loops and specify data layout relationships --- the \skel and \dcs, respectively. We provide heuristics and tactics for extracting these templates from programs written in \lang, an intermediate verification language that powers \sys, and for constructing relational invariants that help the solver performance. 
\item 
Our evaluations with the \sys prototype show that it can successfully verify several complex \mappings for two 
recent open-source 
ML accelerators~\cite{tambe2021flexasr,whatmough2019hlscnn}. Some of these are beyond the reach of existing verification tools/techniques.
\end{itemize}
%
As mentioned earlier, a recent  effort~\cite{melchert2025cgraverif} also considers hardware semantics, but they require information about individual steps in a custom compiler flow, and generate 
restricted classes of reconfigurable accelerators; they do not handle black-box mappings to existing coarse-grained accelerators. 
We also note that our most complex mapping -- linear-layer for the FlexASR accelerator 
-- has 28 loops (many nested, up to maximum depth 7). As far as we know, none of the existing program equivalence techniques/tools (based on either automated reasoning or interactive theorem-proving) has demonstrated successful verification of such complex product programs.

\end{ignore}

\section{Overview of the \sys framework}
\label{s:overview}


In this section we use a motivating example to illustrate the challenges in verifying a \camapping, and show how the key components and steps of \sys enable successful verification.

\subsection{Running Example}
\label{s:overview-example}

\begin{figure}
  \noindent\makebox[\textwidth][c]{%
  \input{figures/flin-mapping-huge}}
  \vspace*{-7mm}
  \caption{Running example: the \mapping for a linear-layer operation in FlexASR.}
  \vspace*{-3mm}
  \label{fig:intro-mapping}
\end{figure}

Consider the \camapping shown in Fig.~\ref{fig:intro-mapping} for a recent open-source
ML accelerator called FlexASR~\cite{tambe2021flexasr}.
This mapping offloads a linear-layer operation to an intrinsic in FlexASR, as used in the 3LA compiler~\cite{huang2024_3la}. 

\myparagraph{\Cspf for linear-layer.} The purple box (upper left) shows the \cspf, in a basic pure-functional tensor language that uses \emph{maps} and \emph{reduces} to perform tensor computations.
These are highly structured loop operations that maintain multidimensional tensor representations for data accesses. In particular, $\csMap{i=0}{n} e$ constructs a tensor by mapping each $i \in \{0, \ldots, n-1\}$ to the result of $e$ at $i$, \IE $[e|_{i=0} \cdots e|_{i=n-1}]$, and  $\csRed{i=0}{n}(s = e_1 \csIn e_2)$ accumulates the value of $s$, starting at $s = e_1$ and updating $s = e_2|_{i, s}$ for each successive value of $i \in {0, \ldots, n-1}$.

\myparagraph{\Aspfue for FlexASR linear-layer.}
The yellow box on the bottom left in Fig.~\ref{fig:intro-mapping} shows the \aspfue for performing a linear-layer using the FlexASR accelerator, where the host processor is an Arm Cortex-A53 (supporting A32 assembly). 
Several MMIO commands configure the accelerator before the final command (emphasized in red) triggers the linear-layer computation within the accelerator hardware.

\subsection{\sys in action}

\begin{ignore}
\begin{figure}[t]
  \includegraphics[width=0.88\columnwidth]{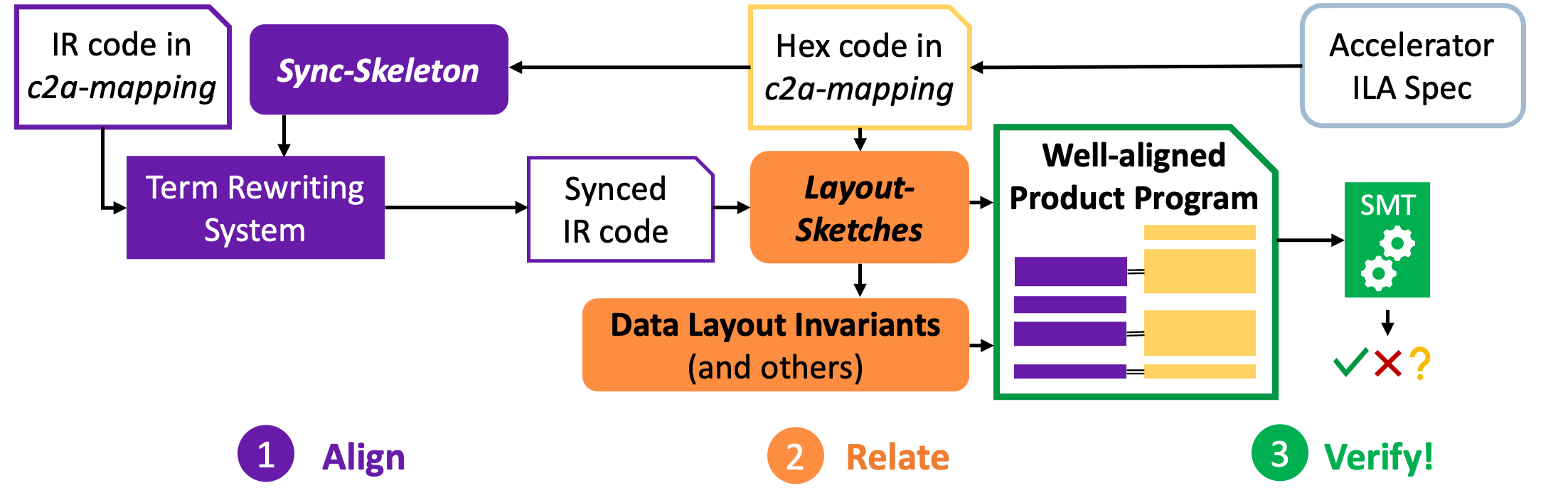}
  \vspace*{-2mm}
  \caption{\sys Framework: Components and Main Steps.}
  \label{fig:methodology}
  \vspace*{-1mm}
\end{figure}
\end{ignore}

We briefly describe the main steps and components of \sys (Fig.~\ref{fig:methodology}), illustrating them using the running example (with pointers to related sections).

\myparagraph{\lang language (\S\ref{s:bolt-hex}).} \sys is powered by the \lang IVL, which provides a common bridge between functional IR code on the one hand with state-based accelerator code that respects the formal hardware semantics (using the ILAs~\cite{huang2018ila}) on the other.
We view \lang as a repurposing of existing HLS-style languages (e.g., C/C++/SystemC), with custom features to support verification of hardware loop nests and complex data layouts.
In particular, \lang supports the development of \skel and \dcs (itself a first-class object),  
by identifying hardware counters and parameters as first-class objects, for reasoning about loop bounds and address arithmetic for data layouts in the \stepT and \stepR steps. 
(In the \stepV step, we substitute parameters with concrete values due to solver limitations with nonlinear arithmetic, \S\ref{s:bolt-verify}.) 
Additionally, \lang is used to represent the product program and supports user annotations of relational invariants that help improve solver performance.  

For the running example,
the yellow box on the right in Fig.~\ref{fig:intro-mapping} shows the \aspf. 
Each \hCall statement (lines 6-10) corresponds one-to-one with an MMIO command, 
after a preamble that defines \emph{parameters} (lines 1-5) that represent key dimensions and sizes for configuring the accelerator.
%
The code block in lines 10-27 shows (a part of) the computation triggered due to the operation 
in line 10.

Note that the FlexASR accelerator code has \emph{28 hardware loops} that implement the linear-layer operation! These are represented in \lang via imperative-style loops, where the \hPfor loops capture 
hardware parallelism due to parallel processing elements (\EG line 11) or single-instruction-multiple-data (SIMD) parallelism (\EG lines 16, 21).
Note also that \lang captures memory accesses and assignments (\EG lines 17, 23-25) that manipulate tensor data in hardware.
\akash{These accesses often involve significant address arithmetic to to optimize tensor data layouts in memory buffers: Line 17 has a simple indirect memory access (due to replicated control state across processing elements) that reads from \texttt{peInOff}, a buffer of address offsets computed in a parent loop. Data layouts are further complicated by overwriting and lazy initialization of buffers, which featured heavily in our other mappings (\S\ref{s:evaluation-key-stats}).}

\myparagraph{Loop alignment via the \Skel (\S\ref{s:bolt-align}).} 
The first main step (\emph{Align}) in \sys is to align the loops on the software side with those on the hardware side. 
This alignment is guided by the \skel, a template extracted from the \aspf. We provide
a \lang-based heuristic procedure for the extraction, such that each loop in the \skel identifies a tensor computation (a map/reduce operation) in the hardware, while loops with only data movement are abstracted away. 
The reason for abstracting away the data movement loops is that they do not appear in the functional \cspf, so there is no need to align them. However, these data movement loops \emph{will} be included at a later stage when the product program is constructed, since they determine tensor data layouts in the hardware buffers.

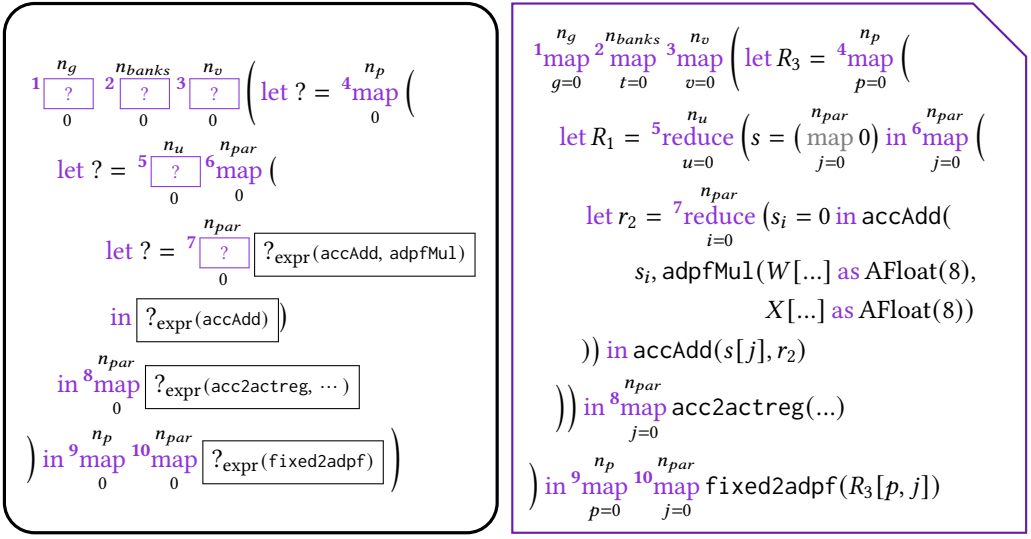
\begin{figure}
  \normalsize
   \definecolor{relateBoxColor}{HTML}{671BA8}

\begin{tikzpicture}
  \node[rectangle,minimum width=0.1\textwidth] (a)
  \bgroup\begin{minipage}{0.45\textwidth}
    \begin{csfrag}
      &\csIter{0}{n_{g}}
        ~\csIter{0}{n_{banks}}
        \csIter{0}{n_v} \bigg(
            \csLet\ ? = \csMap{0}{n_p} \Big( \\
              &\quad
              \csLet\ ? = \csIter{0}{n_u}
                \csMap{0}{n_{par}}\big(\\
              &\qquad\quad
                \mathopen{}\csLet\ ? =
                \csIter{0}{n_{par}}
              \csExpr{\mathtt{accAdd},\ \mathtt{adpfMul}}\\
              &\qquad\quad \csIn \csExpr{\mathtt{accAdd}} 
              \big) \\
              &\quad
              \csIn \csMap{0}{n_{par}}
                \csExpr{\mathtt{acc2actreg},\ \cdots} \\
            &
            \Big)\csIn \csMap{0}{n_p} \csMap{0}{n_{par}}
              \csExpr{\mathtt{fixed2adpf}}~\bigg)
    \end{csfrag}
  \end{minipage}\egroup;

\node[rectangle,minimum width=0.1\textwidth, right=2mm of a] (b)
  \bgroup\begin{minipage}{0.47\textwidth}
      \begin{csfrag}
      &\ \csMap{g=0}{n_{g}}
      \csMap{t=0}{n_{banks}} \csMap{v=0}{n_v} \bigg( \csLet R_3 = \csMap{p=0}{n_p} \Big( && \\
      &\ \quad\csLet R_1 = \csRed{u=0}{n_u} \Big(
        s = \big(\csLoopStyle[gray]{}{\csMapOp}{j=0}{n_{par}} 0\big)
        \csIn \csMap{j=0}{n_{par}} \Big( && \\
        &\ \qquad \csLet r_2 = \csRed{i=0}{n_{par}}
          \big(
            s_i = 0 \csIn
            \mathtt{accAdd}( && \\
              &\ \qquad\qquad \begin{aligned}
                s_i,
                \mathtt{adpfMul}(
                  &W[...] \csAs \tAdpf(8), \\
                  &X[...] \csAs \tAdpf(8)
                )
              \end{aligned} && \\
          &\ \qquad ) \big) \csIn \mathtt{accAdd}(
            s[j], r_2) && \\
      &\ \quad \Big) \Big) \csIn
        \csMap{j=0}{n_{par}}
          \mathtt{acc2actreg}(...) && \\
      &\ \Big)\csIn
        \csMap{p=0}{n_p}
        \csMap{j=0}{n_{par}}
          \mathtt{fixed2adpf}(R_3[p, j]) &&
    \end{csfrag}
  \end{minipage}\egroup;

  \draw [rounded corners=4mm, very thick] (a.west |- b.south) rectangle (a.east |- b.north);
    
  \draw [relateBoxColor, thick]
      (b.south west) 
      -- (b.north west) 
      -- ($(b.north east) - (7mm, 0)$) 
      -- ($(b.north east) - (0, 7mm)$) 
      -- (b.south east) 
      -- cycle; 
\end{tikzpicture}%
   \vspace*{-2mm}
    \caption{Running example: (Left) the \textbf{\skel} (all ten loops) extracted from the \aspf. (Right) the \cspf after rewriting. Loop superscripts identify \textbf{aligned loops} (with same numbering of loops in Figs.~\ref{fig:intro-mapping} and~\ref{fig:overview-dcs}).}
    \label{fig:flin-lss}
    \vspace*{-1mm}
\end{figure}

For the running example, the \aspf (right box of Fig.~\ref{fig:intro-mapping}) clearly identifies the loops in the hardware, including those due to SIMD and PE parallelism.
This loop nest reflects a \emph{fixed} loop tiling and loop ordering in hardware, set by parameters and commands of the given mapping (lines 1-9).
%
Our heuristic procedure enables extracting the \skel shown in the left box of Fig.~\ref{fig:flin-lss}.
The \skel contains 10 loops, corresponding to loops of the \aspf that perform tensor (map or reduce) computations (identified by loop superscripts in Figs.~\ref{fig:intro-mapping} and \ref{fig:flin-lss}) and omits loops that only perform data movement or data initialization. 
Loop types are identified where possible, marking loops as $\csMap{}{}$, $\csRed{}{}$, or either (denoted by $\csIter{}{}$ in Fig~\ref{fig:flin-lss}).

\myparagraph{Rewriting the \cspf for loop alignment (\S\ref{s:bolt-align}).}
The \skel guides a user to perform rewriting on the \cspf to align with the loop nests in hardware. (Since it is easier to rewrite functional programs, we advocate rewriting the \cspf, rather than rewriting the \aspf.) 
We provide a system of rewriting rules, standard and customized (based on \lang features, e.g., parameter substitution rules). A user can employ these rewriting rules to align the loops in the \cspf to those in the \skel. The rewriting and alignment is best-effort, and any un-aligned loops on either side may require additional invariants but do not affect the soundness of verification.

For the running example, the \cspf was transformed through 29 rewriting steps to align with all 10 loops of the \skel.
The resulting \cspf is shown in the right box of Fig.~\ref{fig:flin-lss}, containing the 10 aligned loops (and one unaligned loop that performs zero-initialization).

\myparagraph{Relating tensor correspondences via the \Dcs (\S\ref{s:bolt-relate}).}
After the loops have been aligned, the second main step (\emph{Relate}) in \sys is to relate the tensor data on the software side and the hardware side. 
This is done via a \dcs, a template that 
a user can specify in \lang for each aligned loop. Informally, a \dcs identifies the correspondence between the tensor elements in the \cspf and the buffer elements in the \aspf, respectively, that are updated due to assignments in the aligned loop, assuming some given correspondence between the inputs to the assignment. We provide useful tactics for users to create a \dcs as well as a simple procedure to construct relational invariants from a \dcs, to be added as annotations in the product program. 

\begin{figure}
  \includegraphics[width=.7\columnwidth]{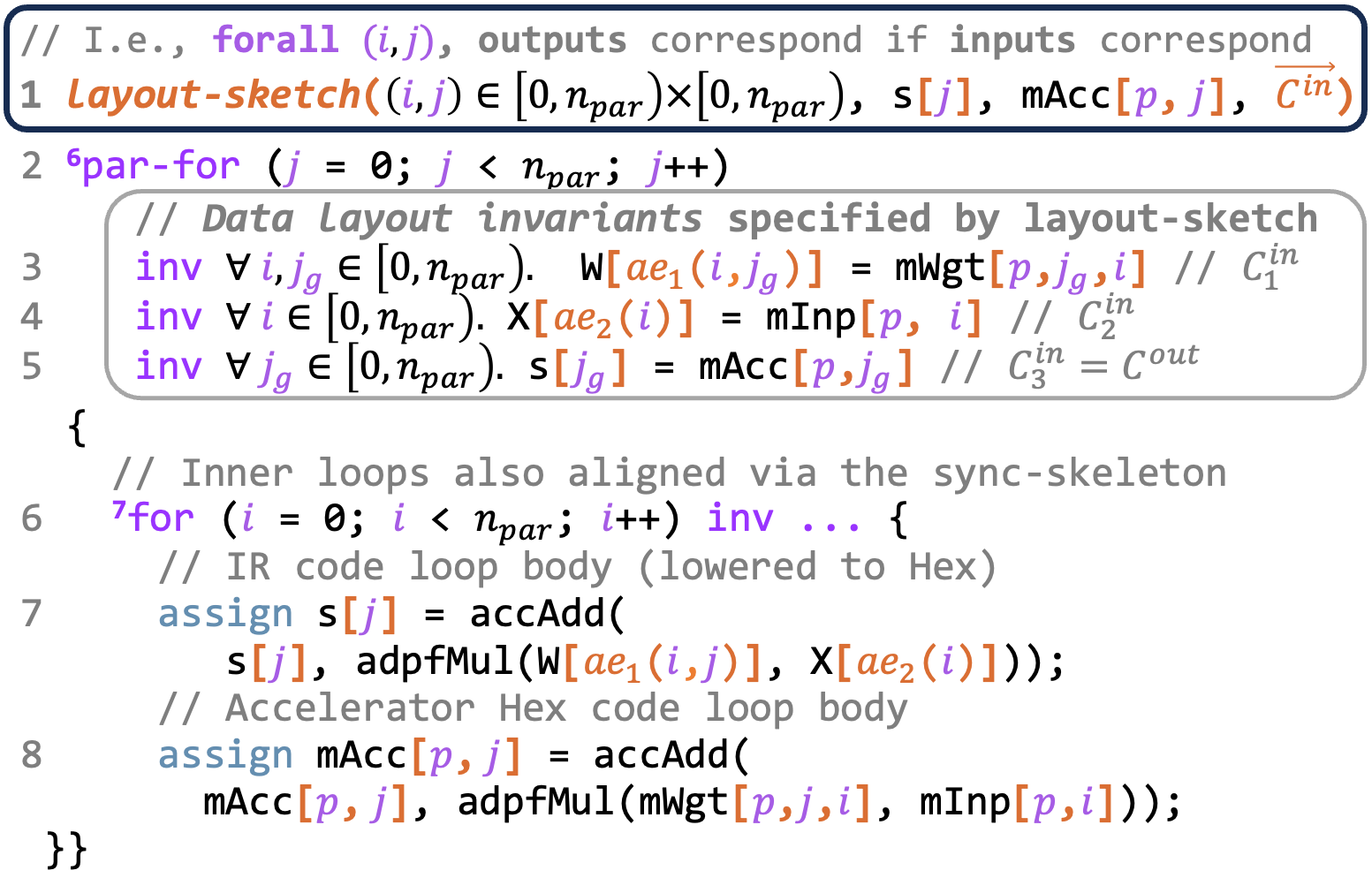}
  \vspace{-3mm}
  \caption{Running example:
  a \dcs and relational invariants for
  an aligned loop (loop \#6, lines 2-8)
  } 
  \label{fig:overview-dcs}
  \vspace{-3mm}
\end{figure}

For the running example, Fig.~\ref{fig:overview-dcs} shows the \dcs (line 1) for an aligned loop. It specifies a correspondence between \cspf tensor element $\hTacc{s}{j}$ with a hardware buffer element $\hTacc{macAcc}{p, j}$, based on a set of input correspondences (elided in the figure).
The resulting relational loop invariants are shown in lines 3-5 (where the $ae$'s represent address arithmetic). 
If any of these invariants is not valid, then the verification fails, thus providing a counterexample for the user to refine the \dcs.

\myparagraph{Verification via Product Programs (\S\ref{s:bolt-verify}).}
The Align and Relate steps enable construction of a \emph{well-aligned} product program in \lang, along with user annotations. In the third step (\emph{Verify}) in \sys, we use standard program verification techniques (e.g., similar to Dafny~\cite{leino2010dafny}) to automatically generate \emph{Verification Conditions (VCs)} based on relational invariants at aligned loop boundaries (and other invariants that a user may provide). These VCs are verified using an off-the-shelf SMT solver (Z3~\cite{demoura2008z3}).

For the running example, we automatically construct a well-aligned product program (along with annotations) from the transformed \cspf (lowered to \lang) and \aspf, and their aligned loops.
Users add the invariants constructed in the \stepR step and any additional invariants necessary for verification.
Note that due to the \stepT step, the loops with tensor operations are already aligned (in a best effort), and due to the \stepR step, relational invariants for tensor data correspondence are defined based on the \dcs.
These two steps significantly help our product program verifier, which generates VCs that are checked by the Z3 SMT solver~\cite{demoura2008z3} (detailed in \S\ref{s:bolt-verify}).
A ``small-world'' model of our running example (with reduced bounds and sizes) was verified in 104s. Verification results for the full model and other case studies are discussed in our evaluations~(\S\ref{s:evaluation}).

\myparagraph{\akash{User effort for tasks in the \sys prototype.}}
In our proof-of-concept prototype, we demonstrate \emph{tractability} of program-equivalence based verification of \camappings, rather than full automation. We briefly summarize the user effort below, where each step largely follows our proposed procedure (setting the stage for future automation).
\begin{enumerate}
\item Align - sync-skeleton extraction — Follow procedure in \S\ref{s:bolt-align-skeleton}, heuristics in \S\ref{s:bolt-align-loop-heuristics}
\item Align - align IR code with sync-skeleton — Follow  procedure in \S\ref{s:bolt-align-rewriting} 
\item Relate - identify layout-sketches — For each loop, follow tactics in \S\ref{s:bolt-relate-sketch-tactics}
\item Relate - specify data layout invariants — Follow procedure in \S\ref{s:bolt-relate-invariants}
\item Verify - construct product program — automated; user adds invariants here 
\item Verify - generate and check verification conditions — automated (using Z3). 
\end{enumerate}

Note that only task (2) can introduce unsoundness, via incorrect rewrite rules --- these can be checked using other tools (e.g., ATL [29]). In particular, sync-skeleton extraction is sound and deterministic, alignment is best-effort, product-program construction is standard and has been shown sound in prior work [9], and the invariants added in Relate and Verify are checked during verification.

\section{\lang: A Verification Language for \underline{H}ardware \underline{Ex}plication}
\label{s:bolt-hex}


\hideableoutline{
\begin{itemize}
  \item We use standard techniques to translate the ILA state transition system into a \lang program.
  \item \lang encourages \emph{parametric} accelerator models and programs for scalability of verification. Parameters can specify bounds in loop nests and dimensions/sizes in data layouts.
  \item \lang bridges the tensor operations of the \cspf and the transition-system loop nests of the accelerator using range-based for loops. \akash{We extend the standard imperative-\hFor construct to identify how a loop has been implemented in hardware, \IE through hardware data-parallelism or hardware registers that act as counters. These annotations enable Align-step heuristics that determine whether a hardware loop should be included as a map/reduce operation in the sync-skeleton.}
  \item \lang provides a standard multidimensional buffer type to describe sets of related registers/buffers. \akashc{Omitted, may depend on Grigory's QI implementation: \akash{\lang pays special attention to buffer accesses that can be factored into \emph{data tilings}, which power a tensor-aware backsubstitution simplifier.}}
  \item \lang does not capture the details of \emph{custom numerical representations} in hardware, treating their operations as uninterpreted functions. \sys checks program equivalence modulo uninterpreted numeric operations, working around the numeric in order to reason about surrounding loops and data.
\end{itemize}
}

We present a new Intermediate Verification Language (IVL)  called \lang 
to reveal the formal hardware semantics 
of an accelerator's intrinsics, for the purpose of proving semantic equivalence between the \cspf and \aspfue. 
The syntax for \lang is shown in Fig.~\ref{fig:hex-syntax}. Wherever possible, we borrow standard constructs from well-known imperative verification languages such as Dafny~\cite{leino2010dafny} \akash{and Why3~\cite{bobot2011why3}}.
Additionally, \lang 
emphasizes hardware loop nests and data layouts 
to support the main steps in the \sys framework as follows:
\begin{itemize}
\item A variant of the \emph{ranged-based} \hFor-loop helps align the accelerator \lang code with the structured loops of an \cspf tensor operation. Additionally, it captures how each loop is implemented in hardware (\EG via SIMD or counter-based iteration), providing heuristics for constructing the \skel in the \stepT step of \sys (\S\ref{s:bolt-align}).
\item The multidimensional buffer type provided in \lang  is the basis for a user to develop a \dcs in the \stepR step in \sys, which relates the data tiling of a tensor in the \cspf with multiple buffers or registers distributed in hardware (\S\ref{s:bolt-relate}).
\item We also use \lang to represent the product program in the \stepV step, including annotations for relational loop invariants (\S\ref{s:bolt-verify}).
\end{itemize}

\begin{figure}
  \centering
  \input{figures/hex-lang-small}
  \vspace*{-.6cm}
  \caption{The syntax of \lang, with terminal symbols for a parameter $n$, variable $x$, and uninterpreted function $f$.}
  \label{fig:hex-syntax}
\end{figure}

\akash{Essentially, a Hex program provides a structured-program view of a transition-system-based hardware semantics.}
The key features of \lang that support modeling the hardware semantics in our context and program equivalence verification are briefly highlighted below. \akash{Note that in principle, an existing verification language like Dafny could be extended with these features and replace Hex in the Bolt framework --- we do not claim any novel contribution in Hex.}

\myparagraph{Parameters for Bounds/Dimensions/Sizes.}
As shown at the top of Fig.~\ref{fig:hex-syntax}, 
\lang supports declaration of parameters, to specify bounds in loop nests and dimensions/sizes in data layouts.
In particular, it supports \emph{application} and \emph{hardware parameters} to capture the dimensions, sizes, and loop bounds, used by the application or fixed by the hardware design, respectively. For parameters that depend on aspects of both sides, \lang supports \emph{derived parameters} that are computed from a combination of hardware and application parameters.

All parameters for our illustrative example are shown in lines 1-2 of the bottom half of Fig.~\ref{fig:intro-mapping}.
The relationships between parameters are explicitly specified using assumptions at the start of a \lang program, as shown in lines 3-5. These relationships are important for manipulating \cspf loop nests in the \stepT step.
Note that while parameters are symbolic in the \stepT and \stepR steps of \sys, we substitute parameters with concrete values in the \stepV step to overcome solver limitations in handling non-linear arithmetic (detailed in \S\ref{s:bolt-verify}).
Thus, while \sys does not produce correctness proofs for parametric models, we encourage using parameters for scalability: a parametric accelerator model can be quickly verified under a ``small-world'' model before 
trying larger bounds and sizes. In future work we plan to explore parametric verification using the Lean interactive theorem prover~\cite{moura2021lean}.


\myparagraph{Capturing the formal hardware semantics.}
\lang provides a \hCall statement to explicate a hardware state update, building on a formal state transition system model of hardware -- \sys uses an ILA model~\cite{huang2018ila}.
In an ILA model, the semantics of a top-level accelerator intrinsic (\EG linear-layer) is captured by an ILA state update, which is defined hierarchically by a sequence (including iterative loops) of hardware state updates. We use standard techniques to translate the ILA transition system into a program in \lang,
where a \hCall in the \lang program corresponds one-to-one with a hierarchically-defined ILA hardware state update.\footnote{\akash{Essentially, we extract (i) parallel state updates, and (ii) ``back-edges’’ in the ILA control flow, and explicate them with Hex par-for and for loops, respectively. Sequential logic is expressed via sequences of \hAssign and \hIf statements.}}
Additionally, \lang makes explicit the loops that occur implicitly in the ILA model due to SIMD- or PE-based parallelism, or via iteration using hardware counters.
Sets of buffers or registers that are used to hold tensor data in the ILA model are represented as multidimensional buffers in \lang.

\myparagraph{Loop Constructs.}
\lang captures tensor map/reduce operations and data movement using range-based for loops (\hFor) and a parallel variant (\hPfor), as shown in Fig.~\ref{fig:hex-syntax}. The \hFor and \hPfor loops use a common syntax, augmented to describe hardware:
Each loop introduces a new counter, $x_i$, that is incremented until it reaches a limit $n$.
If $n$ depends on the application (as either an application parameter or a derived hardware parameter), then the hardware must track $x_i$ via one or more of its registers.
These \emph{hardware counter registers} (in short \emph{hardware counters}), $\overrightarrow{x_{hw}}$ are captured as first-class objects in a $\hWhere-\hBy$ clause,
where each hardware counter $x_{hw} \in \overrightarrow{x_{hw}}$ is related back to the loop counter $x_i$ by a corresponding expression $e \in \overrightarrow{\langle e \rangle}$ that depends only on parameters, loop counters, and constants.
For example, the \hFor loop in Fig.~\ref{fig:intro-mapping}, line 12, has a
${\hWhere-\hBy}$ clause that explicitly associates the loop counter $u$ with the hardware counter $\mathtt{pe\_ctr\_in}[p]$, a PE register.
Like other imperative IVLs,
\lang also supports annotating loops with loop invariants, $\langle \overrightarrow{p} \rangle$, which are checked in the standard assume/assert fashion
during verification.


\myparagraph{Buffers and Tensor Data.}
\lang provides a standard multidimensional \hBuffer type for describing sets of related hardware registers/buffers, \EG the \hBuffer, \texttt{macAcc}, in Fig.~\ref{fig:intro-mapping}, line 23,
corresponds to the accumulators in FlexASR's PEs, where $\hTacc{\mathtt{macAcc}}{p, j}$ denotes the $j$th accumulator of the $p$th PE. Note that as \lang describes physical hardware, all updates to buffers and registers are atomic and globally visible. \sys assumes \lang code is free of data races.

\myparagraph{Custom Hardware Numerics.}
Machine learning accelerators sometimes use \emph{custom numerical representations} to optimize the trade-off between numerical accuracy and speed.
For example, FlexASR uses multiple fixed-point representations of various bitwidths, and a custom numeric called AdaptivFloat~\cite{tambe2020adaptivfloat}.
\lang does not capture the details of these numerics, treating their operations as uninterpreted functions
on the accelerator 
side.
To allow alignment of loops, we introduce the same custom numerics and associated operations on the \cspf side in the \stepT step via \lang-based custom rewrite rules (see next section, \S\ref{s:bolt-align}). This allows verification (in the \stepV step of \sys) to check program equivalence modulo uninterpreted functions.


\section{\stepT step: Aligning the \cspf with loop nests in hardware}
\label{s:bolt-align}

In the \stepT step, users transform the \cspf to match the loop nests of the \aspf, laying the groundwork for the \stepR and \stepV steps that follow.

\subsection{\skeleton}
\label{s:bolt-align-skeleton}
\hideableoutline{
\begin{itemize}
  \item The \skel is a functional program--style sketch of the tensor operations in the \aspf, abstracting away buffers and data-movement loops. The \skel enables use of standard \emph{tensor program rewriting} techniques~\cite{liu2022atl,smith2021glenside} to align the \cspf loop nests with the \skel, and (by extension) with the tensor-computing loop nests of the \aspf.
  \item Table~\ref{tab:skel-patterns} shows the components of a \skel. Loop-free blocks, sequences, and conditionals are respectively mapped 1:1 to the first three rows. We use two heuristics (\S~\ref{s:align-loop-heuristics}) to (i) decide which loops to include in the \skel, and (ii) determine the corresponding map/reduce operation (row 4) or conservatively map a loop to an unknown loop (row 5).
  \item (Uninterpreted) custom numeric operations are retained in the \skel. User-provided rewrites on custom numerics are then used to introduce custom numeric operations in the \cspf, to help align the \cspf.
\end{itemize}
}

We present an abstraction of the \aspf, called a \emph{\skel}, as a template that guides the user for transforming the \cspf.
It is a functional program--style template 
of the tensor operations in the \aspf,
that are comprised of statically bounded \hFor and \hPfor loops.
To extract these tensor computations from their imperative context in the \aspf, the \skel abstracts away:
(i) the data buffers that store tensor data, and (ii) the data-movement loops. 
(Recall that data-movement loops are included in the product programs but do not need to be aligned.)
The \skel enables use of standard \emph{tensor program rewriting} techniques~\cite{liu2022atl,smith2021glenside} to align the \cspf loop nests with the \skel, and (by extension) with the tensor-computing loop nests of the \aspf.

\myparagraph{Components of the \skel.}
A \skel represents a tensor computation with five types of patterns, as shown in Table~\ref{tab:skel-patterns}. These patterns are matched with the \cspf during the rewriting process.

\begin{table}[t]
  \begin{tabular}{p{5em} l p{24em}}
  Name & Pattern & In \cspf, can match any: \\
  \hline \\[-2.5ex]
  \emph{Simple Expr}
  & $\csExpr{ops}$
  & loop-free computation with the operations in $ops$ \\[0.4em]
  \hline
  \emph{Let Assignment}
  & $\csLet\ ? = v \csIn b$
  & $\csLet$ whose value expression matches pattern 
  $v$, and whose body matches pattern $b$.
  Wildcard ``?'' is any variable name. \\
  \hline
  \emph{Conditional}
  & $\mathrm{ite}(c, b_t, b_f)$
  & conditional ite with same condition $c$ and T/F branches matching $b_t$, $b_f$, resp. \\
  \hline
  \emph{Map/Reduce Loop}
  & $\displaystyle\csMap{0}{n} b$, $\displaystyle\csRed{0}{n} b$
  & corresponding map or reduce, resp.,
  with the same loop bound $n$ and body matching
  the pattern $b$
  \\
  \hline
  \emph{Unknown Loop}
  & ${\displaystyle\csIter{0}{n}}\ b$
  & loop with the same loop bound $n$, and
  whose body matches the pattern $b$.
\end{tabular}
  \caption{\Skel components: pure-functional patterns for matching with the \cspf.}
  \label{tab:skel-patterns}
  \vspace{-8mm}
\end{table}

\myparagraph{\akash{Procedure for} \skel extraction.}
The \skel is directly constructed from control flow constructs in the \aspf (see the running example in Figs.~\ref{fig:intro-mapping} and \ref{fig:flin-lss}). Loop-free blocks, sequences of control flow statements, and \hIf{}{} statements are respectively mapped to simple expr, let assignment, and conditional (\texttt{ite}) patterns. A \hFor or \hPfor loop is extracted if it performs a computation on tensor data
, and is then mapped either (a) to the corresponding $\csMap{}{}$/$\csRed{}{}$ loop pattern if the tensor operation can be determined; or {(b)} to an unknown loop, otherwise.
We next describe the heuristics 
to find the corresponding loop pattern for tensor-computing loops, and to identify the pure data-movement loops that are ignored in the \skel.


\subsection{Heuristic procedure for loops in the \skel}
\label{s:bolt-align-loop-heuristics}

\hideableoutline{
\begin{itemize}
  \item Heuristic 1. The \skel should only include loops that perform tensor computations. We first identify \emph{tensor data state} using input tensors and their forward cone-of-influence. A loop performs a tensor computation iff it performs operations on tensor data state.
  \item Heuristic 2. We distinguish map/reduce loops through the presence of extra \emph{accumulator} state needed by reduce loops to hold combined results. We conservatively identify extra accumulator state via a ``data-race freedom''-style check. Loops without accumulators are definitely maps; others can be conservatively marked as unknown loops.
\end{itemize}
}

%

To extract the \skel from the \aspf, we present a procedure composed of two heuristics, which respectively
decide: (1) which \aspf loops to include in the \skel, and (2) which \skel loop pattern to use for a given \hFor loop. These heuristics benefit from \lang loop annotations: hardware-counter annotations on \hFor loops simplify reasoning about hardware control state (by relating it back to loop counters), and \hPfor loops, being parallel, must be $\csMap{}{}$ operations to prevent data races.

\myparagraph{\Skel inclusion heuristic.}
The \skel should only include loops that perform tensor computations, \IE operations on \emph{tensor inputs} or intermediate \emph{tensor data state}. In the \aspf, we know which buffers contain tensor inputs through the initial correspondence between the \cspf and \aspf at the start of the \camapping. We conservatively identify intermediate tensor data state in the \aspf as the forward cone-of-influence of these buffers. 
A loop should be included in the \skel if it performs any operation other than a load/store on tensor inputs/data state; all others perform data movement only and should be omitted.

\myparagraph{Hardware-map heuristic.}
We identify that a \aspf loop corresponds to a map operation by checking for the \emph{absence} of any \emph{accumulator state} that is used by reduce operations to hold intermediate combined results. Since an accumulator is reused across multiple iterations of a loop, we conservatively test for accumulators with an SMT check for read-write conflicts:
given a loop, we check whether the same register or buffer element $\hTacc{B}{\overrightarrow{addr}}$ is written in one iteration and read/written in another.
Formally, let $R_B$ and $W_B$ be sets of formulas denoting the read addresses and write addresses of buffer $B$ inside the loop. Let $L_{pars}$ be the counters of parent loops, and $x_{l1}, x_{l2}$ and $L_{child1}, L_{child2}$ be two copies of the loop counter and counters of child loops respectively. Buffer $B$ contains accumulator state if the following query is satisfiable:
\[
    x_{l1} \neq x_{l2}
    \land \bigvee_{\vec{w} \in W_B} \overrightarrow{addr} = \vec{w}(L_{pars}, x_{l1}, L_{child1})
    \land \bigvee_{\vec{o} \in W_B \cup R_B} \overrightarrow{addr} = \vec{o}(L_{pars}, x_{l2}, L_{child2}).
\]
If every register or buffer element written in a loop does not contain accumulator state, the loop is definitely a $\csMap{}{}$; otherwise, it is soundly mapped to an unknown loop in the \skel.

\myparagraph{\akash{Effectiveness of heuristics.}} The inclusion heuristic is effective when the hardware performs only the tensor computation being specified, and less so when there are other unrelated tensor computations co-occurring (rare -- we did not see examples of this in our evaluations). 
The hardware-map is effective when there is low overwriting during map operations, \IE when an input tensor is not in-place replaced by an output tensor (of different dimensions). Note that reduce operations always have overwriting, so loops with complex data layouts such as non-stationary output accumulation are correctly categorized by the heuristic as unknown loops in the \skel.

\subsection{Rewriting the \cspf}
\label{s:bolt-align-rewriting}

\hideableoutline{
\begin{itemize}
  \item Table~\ref{tab:rewrite-rules} shows the categories of rewrite rules that we provide in \sys. \sys combines \emph{arithmetic} and \emph{structural} rewrite rules with three new categories of rewrite rules that are guided by the hardware: \emph{parameter substitution}, \emph{custom numeric}, and \emph{type casting}.
  \item Although we aim to align all loops, incompleteness of the rewriting rules may result in a \emph{best-effort} alignment with some unmatched loops on either side. Unmatched loops may require additional invariants and may affect the solver performance in the Verify step, but do not affect the \emph{soundness} of our verification.
\end{itemize}
}

After a \skel has been extracted, a user aligns the \cspf with the \skel by using rewriting techniques that replace parts of IR code with equivalent alternatives, based on a set of source-to-source \emph{rewrite rules}.
Prior work~\cite{liu2022atl,smith2021glenside} 
has developed \emph{structural} rewrite rules for exploring loop nests via transformations like loop tiling and loop reordering. \sys combines structural rules with four other categories of rewrite rules---shown in Table~\ref{tab:rewrite-rules}---to achieve cheap and expressive rewriting \emph{via hardware guidance}.
\sys depends on the correctness of these (simple) rewrite rules.
For our running example, rules from these five categories are used to match all 10 \skel loops.
We highlight three \emph{new} categories of hardware-guided rewrite rules. 

\begin{table}
  \noindent\makebox[\textwidth][c]{%
  \begin{minipage}{0.47\columnwidth}
  \begin{tabular}{l}
    \hline \\[-2ex]
    \multicolumn{1}{l}{\emph{Arithmetic (A)}} \\
    $(a + b) * c \Leftrightarrow a * c + b * c,$ \quad $a - a \Leftrightarrow 0$
    \hfill{} \\
    \hline \\[-2ex]
    \multicolumn{1}{l}{\emph{Structural (S)}} \\
    {$\!\begin{aligned}
    &\csMap{i=0}{n_1 * n_2} e_1 \Leftrightarrow
    \csMap{i=0}{n_1}\csMap{j=0}{n_2} e_1[i \mapsto n_1 * i + j]
    \end{aligned}$} \hfill{} \\
    \hline \\[-2ex]
    \multicolumn{1}{l}{\emph{Parameter substitution (P)}} \\
    {$\!\begin{aligned}
      & n_t \Leftrightarrow n_g * n_{banks}, \;
      & n_i \Leftrightarrow n_u * n_{par}
    \end{aligned}$}
    \hfill{} \\
    \hline \\[-2ex]
    \multicolumn{1}{l}{\emph{Custom numeric (N)}} \\
    {$\!\begin{aligned}
    &(e_1: \R) * (e_2: \R) \Leftrightarrow \big(\mathtt{afloatMul}(\\
    &\quad e_1 \csAs \tAdpf(n),
    e_2 \csAs \tAdpf(n)) \csAs \R\big)
    \end{aligned}$} \\
    \hline \\[-2ex]
    \multicolumn{1}{l}{\emph{Type casting (T)}} \\\
    {$((e_1 \csAs \tau_1) \csAs \tau_2) \Leftrightarrow (e_1 \csAs \tau_2)$}
    \hfill{}\\
    \hline 
\end{tabular}
  \caption{Rewrite rules: Categories of rewrite rules for aligning the \cspf with the \skel.}
  \label{tab:rewrite-rules}
  \end{minipage}\hspace{0.5em}%
  \begin{minipage}{0.52\textwidth}
  \begin{tabular}{r p{10em} l r}
  \multirow{2}{*}{\#} & \multirow{2}{*}{Rewrite rule} & \multirow{2}{*}{Cat.} & \multicolumn{1}{p{1.3cm}}{\multirow{2}{1.3cm}{Loops matched}} \\ \\
  \hline
  1 & let-substitute & S & 0/10 \\
  2 & $n_t \Leftrightarrow n_g * n_{banks}$ & P & 0/10 \\
  3 & map-tiling & S & 2/10 \\
  4 & $n_o \Leftrightarrow n_{vp} * n_p$ & P & 2/10 \\
  5 & $n_{vp} \Leftrightarrow n_v * n_{par}$ & P & 2/10 \\
  6 & associativity & A & 2/10 \\
  ... & various & A,S,P & 7/10 \\
  15 & mul-adpfloat & N & 7/10 \\
  16 & acc-add & N & 7/10 \\
  17 & cast-recast & T & 7/10 \\
  ... & various & all & 8/10 \\
  29 & map-let-reordering & S & 10/10
\end{tabular}
  \caption{\Cspf alignment: The sequence of rewrite rules used in the running example \stepT step.
  Letters in column 'Cat.' correspond to categories in Table~\ref{tab:rewrite-rules}.}
  \label{tab:flin-cspf-rewriting}%
  \end{minipage}}
  \vspace{-8mm}
\end{table}


\myparagraph*{Parameter substitution rules} use known relationships between application and hardware parameters to guide structural rewrites towards loop nests that reflect hardware bounds and sizes. These are derived from the preamble of the \aspf (see the bottom half of Fig.~\ref{fig:intro-mapping}, lines 3-5).

\myparagraph*{Custom numeric rules} are user-specified rules for manipulating (uninterpreted) operations in a custom numeric, for either (i) converting to the custom numeric from real arithmetic (shown in Table~\ref{tab:rewrite-rules}), or (ii) converting between custom numeric operations. 
A user can guide their application using the \skel, by preferring rewrites that result in one or more of the required operations in a simple expr, $\csExpr{ops}$.

\myparagraph*{Type casting rules} are generic rules that simplify and propagate type casts to better reflect numeric conversions in hardware. The type-casting rule in Table~\ref{tab:rewrite-rules} simplifies nested casts. \\

For our running example, Table~\ref{tab:flin-cspf-rewriting} lists the sequence of 29 rewrites that a user performs to align the \cspf with the \skel.
Combining parameter substitution rules with standard structural and arithmetic rewrites enables matching 7 of 10 loops of the \skel, \IE all loops except 4, 9, and 10.
By further adding custom numeric and standard type casting rules, 
the \cspf is transformed to fully match the \skel, as seen in the last line of Table~\ref{tab:flin-cspf-rewriting}.
Custom numeric rewrite rules are specific to each accelerator design, such as the rules used in our running example (one of which is shown in Table~\ref{tab:rewrite-rules}).
However we found the majority of rewrite rules used (structural, arithmetic, type casting) are standard across benchmarks (\S\ref{s:evaluation-key-stats}), and parameter substitution rules are extracted from the preamble of the accelerator fragment.
 

\myparagraph{Best-effort Matching.}
For all the case studies we have evaluated so far (\S\ref{s:evaluation}), the above extraction heuristics and rewrite rules were effective in matching completely, or most of the loops in the \skel.
Although we aim to align \emph{all} \cspf loops to those in the \skel, due to incompleteness of the rewriting rules in practice, this step may result in a \emph{best-effort} alignment that may leave some unmatched loops on either side, \IE some loops may remain \emph{unaligned}. We note that such unaligned loops may require additional invariants and may affect the solver performance in the Verify step (\IE it may fail or take more time). However, they do not affect the \emph{soundness} of verification.
In the future, we plan to investigate equality saturation frameworks like egg~\cite{willsey2021egg} for exploring a larger/complete space of \cspf transformations.

\section{\stepR step: Relating the tensor data layouts}
\label{s:bolt-relate}

After the loops are potentially aligned in the \cspf and \aspf due the \stepT step, the next step in \sys -- 
the \stepR step -- 
is to capture tensor data correspondences between the \cspf and \aspf as relational loop invariants.
\akash{}

\subsection{\dcsfull}
\label{s:bolt-relate-dcs}
\hideableoutline{
\begin{itemize}
  \item After the \stepT step, the loop nests of the transformed \cspf and \aspf have many synchronized loops, and some loops on either side may be unsynchronized.
For each synchronized loop, a \dcs enables us to construct \emph{relational data layout invariants} that are customized to 
the $\csMap{}{} / \csRed{}{}$ operations on the compiler side, and any buffer-overwriting on the accelerator side.
  \item A \dcs for a synchronized loop states that some tensor element $\hTacc{T}{...}$ corresponds to the buffer element $\hTacc{B}{hwae}$ in every loop iteration, assuming a given correspondence between the reads that these assignments depend on. \akashc{Formalization goes here.} Terminology: output/input correspondence, quantified range, access, hardware access.
  \item For unsynchronized loops, such as those performing data initialization or data movement, we can either (i) treat them as a synchronized loop (one side is constant), or (ii) defer handling the loop until the Verify step, where we will manually provide, infer, \akash{or synthesize} additional loop invariants.
\end{itemize}}

For each aligned loop in the transformed \cspf and the \aspf, we propose to develop a \emph{\dcsfull}, 
that describes tensor data correspondences for assignments in the loop bodies. 
Crucially, \dcss bridge the gap between \emph{pure-functional} tensor computations in the \cspf and \emph{imperative} buffer updates in the \aspf.
We provide several tactics for a user to specify a \dcs in \lang, and  
to construct \emph{relational data layout invariants} from a \dcs.

\myparagraph{Aligned loops.} Intuitively, a \dcs for an aligned loop takes an assignment from each side (\EG an \cspf ${\csLet{T} = ...}$, and an \aspf $\hAssign \hTacc{B}{hwae} = ...$ ), and states that some tensor element $\hTacc{T}{...}$ corresponds to the buffer element $\hTacc{B}{hwae}$ in every loop iteration, assuming a given correspondence between the reads that these assignments depend on.

More formally, a \dcs is defined by a tuple 
$(R, \hTacc{T}{ae}, \hTacc{B}{hwae}, C^{in})$, 
where:
\begin{itemize}
  \item $R = \{\vec{i} \in [\vec{0}, \vec{n})\}$ defines a quantified range
  over which the input/output correspondence holds, based on $\vec{i}$ and $\vec{n}$, which are, respectively, the loop counters and loop bounds of the loop and its children loops.
  \item $\hTacc{T}{ae}$ is the LHS term of an \cspf assignment, which updates a tensor $T$ at the multidimensional index $ae$. We call $ae$ an \emph{access}, and express $ae$ in terms of $\vec{i}$, parent loop counters, and parameters
  (based on the $\csMap{}{}$ /$\csRed{}{}$ operations in the \cspf).
  \item $\hTacc{B}{hwae}$ is the LHS term of an $\hAssign$ statement in the \aspf, which updates a buffer $B$ at the address $hwae$. We call $hwae$ a \emph{hardware access}.
  Hardware accesses are written in terms of offsets and hardware counters; we lift them to expressions over loop counters and parameters by back-substitution in the loop body.
  \item $C^{in}$ is a set of input correspondences, where each element $C^{in}_{k} = (\hTacc{T_k}{{ae}_k}, \hTacc{B_k}{{hwae}_k}, \mathtt{isOwr})$ denotes a correspondence $\hTacc{T_k}{{ae}_k} = \hTacc{B}{{hwae}_k}$ in the \cspf and \aspf fragments, respectively. The additional flag \texttt{isOwr} is $\top$ (\IE true) in the case where the \aspf overwrites parts of $\hTacc{B_k}{{hwae}_k}$ with a different tensor's data during the loop iterations. This case occurs in one of our FlexASR case studies (pooling,~\S\ref{s:evaluation}).
\end{itemize}


\myparagraph{\Dcs example.} For our running example (\S\ref{s:overview}), consider a \dcs for the aligned loop over $j$ in Fig.~\ref{fig:overview-dcs} (line 1). It is defined as a tuple $(R, \hTacc{T}{ae}, \hTacc{B}{hwae}, \{C^{in}_1, C^{in}_2, C^{in}_3\}),$ where:
\begin{itemize}
  \item $R = \{(i, j) \in [0, n_{par})\times[0, n_{par})\}$,
  \item $\hTacc{T}{ae} = \hTacc{s}{j}$,
  \item $\hTacc{B}{hwae} = \hTacc{\mathtt{macAcc}}{p, j}$, where $p$ is the loop counter of a parent loop.
  \item $C^{in}_1 = (\hTacc{s}{j}, \hTacc{\mathtt{macAcc}}{p, j}, \bot)$ relates both sides' accumulators at the start of the iteration,
  \item $C^{in}_2 = (\hTacc{W}{ae_1(i, j)}, \hTacc{\mathtt{macWeights}}{p, i, j}, \bot)$
  \item $C^{in}_3 = (\hTacc{X}{ae_2(i)}, \hTacc{\mathtt{macInputs}}{p, i}, \bot)$
\end{itemize}

\myparagraph{Unaligned loops.}
Loops that perform data initialization or data movement, or are unmatched tensor computation--performing loops, may be unaligned at the end of the \stepT step. We have two options for handling them:
\begin{itemize}
    \item \emph{Treat as an aligned loop}: define a \dcs for the unaligned loop where the outputs on one side are constants in the case of initialization, or the same as the inputs in the case of data movement.
  \item \emph{Treat as a single-side loop}: provide invariants manually or unroll the loop during verification. 
\end{itemize}

\subsection{Tactics for developing a \dcs}
\label{s:bolt-relate-sketch-tactics}

We briefly outline some tactics users can follow when developing a \dcs. We mainly consider the important case of \dcss for aligned loops, and touch on constructing a \dcs for data-movement/initialization at the end. \akash{These tactics are most effective when sync-skeleton loops are all aligned and the two sides are mostly in lockstep (aside from data movement).}

\myparagraph{Tactic 1 (Capture lockstep behavior).}
A \dcs is designed to help users specify data layout loop invariants when each side computes the same outputs from the same inputs (with potentially different data layout) \emph{in lockstep}.
For loops whose bodies or inner loop nests may not be in lockstep (\EG they are unaligned despite best-effort matching), a \dcs may be harder to develop, or users may need to provide additional invariants in the \stepV step.

\myparagraph{Tactic 2 (\Cspf loops identify tensor data state).} Every loop in \cspf constructs a new tensor representing the result of the given $\csMap{}{}/\csRed{}{}$ operation based on the inputs and previously constructed tensors. 
The loop should have a \dcs that relates the result tensor to a buffer in the \aspf, with an input correspondence for each access to an input/previous tensor in the operation. 
For each access, 
users must identify the corresponding hardware access on the \aspf side, \IE a buffer write/read somewhere within the loop body.
We plan to automate this process through dataflow analysis in future work.

\myparagraph{Tactic 3 (Copy from nearby/inner aligned loops).} 
The \dcs of a pair of aligned loops can often reuse the quantified range and correspondences of nearby/inner aligned loops, after quantifying over the outer loop counter. We can reuse the output correspondence because the inner loop nests on both sides are still constructing the same \cspf result tensor. We can reuse input correspondences because the input and previous tensors are immutable on the \cspf side (although we may need to capture overwriting within the hardware). 

\myparagraph{Tactic 4 (Data movement / zero-initialization follow standard patterns).}
For both a data-movement loop (which can only be in the \aspf) and a zero-initialization loop, we do not have an aligned loop on the other side.
However, in the case of data-movement, the source hardware access is related by some \dcs (or initial assumption) to some \cspf tensor. Users can relate the destination hardware access to the same tensor by reasoning about address arithmetic.
Zero initialization is even easier, as users only need to specify the order in which tensor/buffer elements are zero'ed, \IE quantifying over the loop counters of loops surrounding the zero-initializing assignment.


\subsection{\akash{Procedure for specifying} data layout invariants from a \dcs}
\label{s:bolt-relate-invariants}
\hideableoutline{
\begin{itemize}
  \item Each output/input correspondence of a \dcs corresponds 1:1 with a data layout invariant.
  \item We relate individual tensor/buffer elements on all iterations of a loop by quantifying over the \emph{shared} loop counters of both sides.
  \item Loop invariant guards can be determined based on the map/reduce nature of the \cspf loop and any input overwriting in the \aspf loop.
\end{itemize}}

Given a \dcs $(R, \hTacc{T}{ae}, \hTacc{B}{hwae}, C^{in})$ for a loop $l$, we provide a procedure to construct relational loop invariants that describe the output correspondence between $\hTacc{T}{ae}$ and $\hTacc{B}{hwae}$ and input correspondences in $C^{in}$.
For each output/input correspondence, a \emph{data layout invariant} takes the following form:
$$\forall \vec{i} \in R. \ g(\vec{i}, i_{cur}) \Rightarrow \hTacc{T}{ae(\vec{i})} = \hTacc{B}{hwae(\vec{i})},$$
stating that the $\vec{i}$th tensor elements on both sides are equal (\IE $T[ae(\vec{i})] = B[hwae(\vec{i})]$) for all $\vec{i} \in R$.
The optional guard $g$ captures incremental progress (or regress) of the loop, essentially considering a subset of $R$ based on the current iteration $i_{cur}$.

\myparagraph{Identifying corresponding tensor elements via (shared) loop counters.}
The data layout invariant relies on identifying the corresponding tensor elements $T[ae(\vec{i})] = B[hwae(\vec{i})]$ for $\vec{i} \in R$. This requires defining the range $R$ and accesses $ae, hwae$ in terms of the same variables $\vec{i}$.
This requirement can be fulfilled by defining $R$ and the accesses in terms of shared loop counters (and parameters and constants) using the \dcs.
This underscores the benefit of the \stepT step, which provides shared loop counters by synchronizing the loop nests on both sides.

\myparagraph{Determining guards from loop context.}
For an output correspondence, the choice of guard $g$ depends on what tensor-computation is being performed in the loop $l$. No guard is necessary ($g = \top$) when $T$ is the \emph{accumulator} of a $\csRed{}{}$ loop ($l$ itself or a parent). Otherwise, $T$ is the output of a $\csMap{}{}$ operation, so $g$ states that the correspondence holds for elements up to the current loop iteration, $i_{cur}$, via the guard $i_l < i_{cur}$, where $i_l \in \vec{i}$ corresponds to $l$'s loop counter.

For an input correspondence $(\hTacc{T_k}{{ae}_k}, \hTacc{B_k}{hwae}_k, \mathtt{isOwr}_k)$, a guard is necessary if the \aspf's buffer is being overwritten (\IE \texttt{isOwr}). In this case, a guard $i_l \geq i_{cur}$ is used to state that the input correspondence holds for the inputs of remaining loop iterations.

\section{\stepV step: Verifying the product program}
\label{s:bolt-verify}



\hideableoutline{
\begin{itemize}
  \item Preamble: background on product program verification and VCs, with manually annotated invariants.
  \item Challenge: VCs from a \lang program typically have assumptions and assertions with quantifiers, \EG from \dcs-based relational invariants. Standard SMT-based approaches (e.g., Z3/Spacer) struggle with solving the resulting problems.
  \item Two approaches to eliminating quantifiers: (1) bounded flattening, remove the $\forall$ quantifiers by flattening in the data layout invariants when used as assumptions --- not scalable, and (2) layout-driven quantifier instantiation.
  \item \lang parameters, if left symbolic, can cause VCs to contain non-linear arithmetic formulae (for address arithmetic). We substitute all parameters with concrete values. Handling non-linear arithmetic is future work.
  \item TODO: Correctness claim
  \item \akash{We also explored the use of Constrained Horn Clause (CHC) solving for program verification, \EG Z3/Spacer~\cite{komuravelli2016spacer,gurfinkel2021chcuf}, where the CHC solver can infer the required invariants.}
\end{itemize}}

The last step in \sys constructs 
a well-aligned product program in \lang from 
the transformed \cspf and the \aspf, along with annotations of the relational data layout invariants. We automate this construction using standard techniques~\cite{barthe2011relational,angelis2016relationalverif}.
%
\sys then uses an automated procedure for verifying the product program 
based on generating Verification Conditions (VCs), using annotations at loop boundaries (in the style of Dafny~\cite{leino2010dafny}), and checking their validity using an SMT solver (Z3~\cite{demoura2008z3}).
\akash{Our SMT formulas contain theories of bitvectors, uninterpreted functions, arrays, and linear integer arithmetic.}
We review the VC-generation procedure and highlight some design choices below.

\begin{figure*}
  \includegraphics[width=\textwidth]{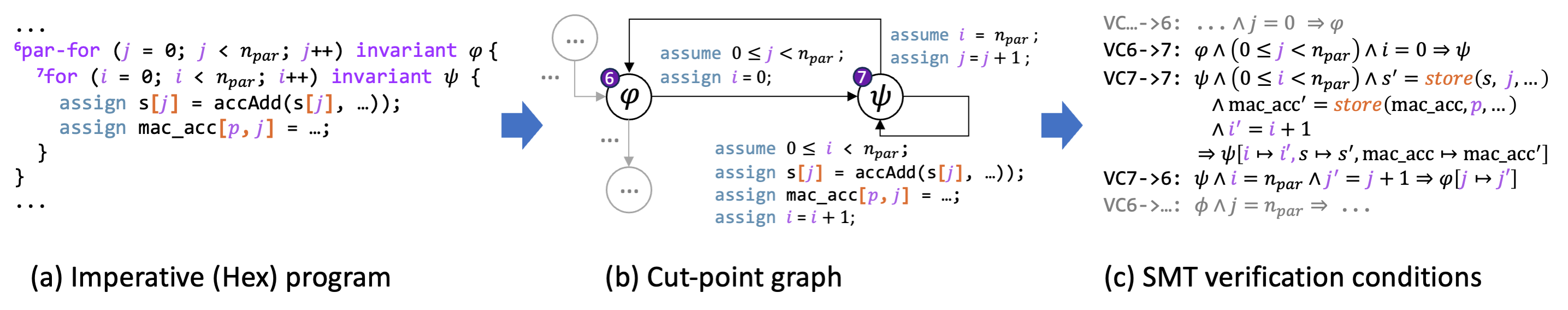}
  \caption{\sys uses standard techniques to generate SMT verification conditions from the \tpp.}
  \label{fig:smt-encoding}
\end{figure*}

\myparagraph{Background: VC generation from product programs.}
\akash{Fig.~\ref{fig:smt-encoding} depicts the standard procedure we use to generate verification conditions from a \lang product program.
Essentially, the (product) program is ``cut'' at its loops
to create a \emph{cut-point graph}, a graph whose nodes are loop heads (and program start/end points),
and whose edges are loop-free program segments. Each node is annotated with a Hoare-style specification, and each edge
is used for generating a VC that must be proved valid via an SMT solver.
If the SMT solver can prove the validity of every VC, then
we have successfully proved the equivalence of the \cspf and \aspf in the product program.
If a VC is invalid, either the assumed annotation is too weak, or the asserted annotation is too strong, or the program contains a bug.}

\myparagraph{Avoiding non-linear arithmetic.}
The VCs generated for our product programs often include address arithmetic with products of parameters.
If these parameters are symbolic, the VCs have
non-linear arithmetic formulae, which are notoriously difficult to solve with standard SMT solvers.
Our verifier avoids solver limitation (for now) by \emph{substituting parameters with concrete values.}
Note that with advances in SMT solvers, or by exploring use of interactive theorem provers like Lean~\cite{moura2021lean}, we hope to support parametric verification in the future, such that program equivalence is verified for all sizes/dimensions represented by the parameters.

\myparagraph{Handling quantifiers.}
VCs from a \lang program typically have assumptions and assertions with quantifiers, \EG from \dcs-based relational invariants. SMT solvers often struggle to reason about quantifiers.
Our strategy
is to remove the universal quantifiers by bounded flattening~\cite{mccune2001mace} (up to constant bounds in the concrete parameters) in the data layout invariants when used as assumptions.
%

\subsection{Categories of useful invariants}

\hideableoutline{
\begin{itemize}
  \item Table~\ref{tab:invariant-categories} lists categories of loop invariants used by our verifier, along with examples.
  \item \akash{We used some simple static analyses to automatically derive many (but not all) required invariants in these categories.} \akashc{Static analyses have not been instrumented to measure how many invariants were found, nor how many were actually useful. They may also be subsumed by tensor-aware backsubstitution.}
  \item \akash{In our CHC-based verification, we synthesized loop bound and WORM invariants based on simulation traces.}
\end{itemize}}

We list the categories of loop invariants used in 
\sys in Table~\ref{tab:invariant-categories}, along with examples. These categories were sufficient in all our evaluations (\S\ref{s:evaluation}).
In addition to data layout invariants (guided by the \dcs, \S\ref{s:bolt-relate}), verification typically needs invariants on loop bounds, hardware counters and registers, briefly described below.
Our verifier in \sys uses simple static analyses to derive many (but not all) required invariants in these categories.

\begin{table}
  \begin{tabular}{p{7.6em}|l} 
  Invariant category & Example \\ 
  \hline
  Data layout &
  $\hLmap t_g \in [0, n_{banks}),
    v \in [0, n_v).$ 
  $(t_g < t \implies$ 
  ${\hTacc{X}{t_g \hTaccSep v} = \hTacc{\mathtt{buf}}{c_1 t_g + c_2 v}})$
  \\
  Loop bound & $0 \leq u < n_u$
  \\
  Hardware-counter &
  $\hForall p \in [0, n_p). {\hTacc{\mathtt{ctr}}{p} = ite(u < n_u, u, 0)}$
  \\
  WORM &
  $\mathtt{addrOff} = \mathtt{BASE\_OUT}\,$
 ${+\ n_p * \mathtt{ctrTsg}}$
\end{tabular}
  \caption{Verify step: Categories of Invariants} 
  \label{tab:invariant-categories}
  \vspace{-.6cm}
\end{table}

\myparagraph{Loop bound invariants.}
The \hFor and \hPfor in \lang are standardized to start at zero and increment at the end of the loop up to some parameter loop bound. This is captured as a loop bound invariant.

\myparagraph{Hardware-counter invariants.}
To analyze hardware address arithmetic, 
it is useful to have invariants for the hardware counters identified in the \hWhere-\hBy clauses of a \hFor loop.
For example, the hardware-counter invariant in Table~\ref{tab:invariant-categories} shows a hardware counter that resets on loop end.


\myparagraph{WORM invariants.}
Hardware designs commonly save 
(computed) address offsets, and 
configuration information in hardware registers for reuse. 
We have found that capturing this \emph{write-once-read-many (WORM)}
pattern as an invariant is useful, especially when the written value is read much later in another loop.

\section{Case Studies and Evaluation}
\label{s:evaluation}
\hideableoutline{
\begin{itemize}
  \item We have developed a proof-of-concept prototype for the \sys/\lang framework and evaluated it for verifying some \camappings for two state-of-the-art ML accelerators, FlexASR and HLSCNN.
\end{itemize}}

We have developed a proof-of-concept prototype for the \sys/\lang framework and evaluated it for verifying  \camappings for two recent open-source 
ML accelerators described below.
We report some key statistics (Table~\ref{tab:key-stats}) for the three main steps (Align, Relate, and Verify) of \sys, and the verification results (Table~\ref{tab:verif-time}) for the product programs using the Z3 SMT solver~\cite{demoura2008z3}.

\subsection{Case studies: Compiler-to-Accelerator Mappings}
\label{s:evaluation-mappings-list}

For our evaluation we used accelerators with available formal hardware semantics. To the best of our knowledge, thus far, only the 3LA compiler framework~\cite{huang2024_3la} provides open-source hardware semantics for ML accelerators. It does so for three accelerators -- FlexASR, HLSCNN, and VTA\footnote{ILAs models for all three accelerators are available at \href{https://github.com/PrincetonUniversity/IMDb/}{https://github.com/PrincetonUniversity/IMDb/}}. It used \camappings for code generation\footnote{Code available at \href{https://github.com/uwsampl/3la-evaluation}{https://github.com/uwsampl/3la-evaluation}}, but did not address proving their correctness. Of these accelerators, VTA has fine-grained mappings, which is not the focus of this work, and thus not considered. 
We consider the other two accelerators, and their mappings (described below) are of significant complexity; verifying their correctness provides a proof of concept of \sys's capabilities.
(We will open-source the mappings used in our evaluations, which we converted from the original DSLs used in 3LA to IR and \lang code for verification.)

%
%

\myparagraph{FlexASR Accelerator.} FlexASR~\cite{tambe2021flexasr} supports various operations used for automatic speech recognition, including linear-layer, pooling, etc. Some operations
are parallelized across multiple PEs 
and others are performed directly using data in a global buffer. 
We verify the linear-layer and pooling as representative mappings in these two categories, respectively.

\begin{itemize}
\item{FlexASR Pooling (FP).} FlexASR supports three pooling operations, called max-, add-, and mean-pooling.
  3LA contained \camappings for the first two; we constructed a similar mapping for add-pooling.
  We used (nearly) the same \stepT and \stepR steps on all three (modulo renaming some variables, loops, etc.), so we describe them together in Table~\ref{tab:key-stats}. However, in the \stepV step, we check each mapping individually to prove its correctness.

\item{FlexASR Linear-layer (FLL).} FlexASR linear-layer has been detailed as the running example, here we focus on its role as a challenge case study: it has the largest mapping, with significant hardware data-parallelism and data-movement.
\end{itemize}

\myparagraph{HLSCNN Accelerator.} HLSCNN is a convolutional-layer accelerator~\cite{whatmough2019hlscnn} that supports the following operations: 
\begin{itemize}
\item 
HLSCNN 2D Convolution (C2D): HLSCNN supports a 2D convolution operation that has variable kernel sizes, and supports batch computation (over filters and channels).
 
\item HLSCNN C2D Accumulative (C2D+): When verifying C2D, we noted a hardware flag---unused by the 3LA mapping---for accumulating outputs onto previous results.
As the flag significantly changes data layout correspondences, we created and verified C2D+, a mapping with the flag enabled.
\end{itemize}


\hideableoutline{
\begin{itemize}
  \item FlexASR Pooling (FP) --- split into 3 mappings for the Verify step to prove correctness of each supported pooling operation (max-, add-, mean).
  \item FlexASR Linear-layer (FLL) --- largest mapping, with significant hardware data-parallelism and data-movement.
  \item HLSCNN 2D Convolution (C2D)
  \item HLSCNN C2D with Accumulation Flag (C2D+) --- When verifying C2D, we noted a hardware flag---unused by the 3LA mapping---for accumulating outputs onto previous results.
As the flag significantly changes data correspondences, we created and verified C2D+, a mapping with the flag enabled.
\end{itemize}}

\begin{ignore}
\myparagraph{FlexASR Pooling (FP).} FlexASR supports three pooling operations, called max-, add-, and mean-pooling.
  3LA contained \camappings for max-pool and mean-pool; we constructed a similar mapping for add-pooling.
  We used (nearly) the same \stepT and \stepR steps on all three (modulo renaming some variables, loops, etc.), so we describe them together in Table~\ref{tab:key-stats}. However, in the \stepV step, we check each mapping individually to prove its correctness.

\myparagraph{FlexASR Linear-layer (FLL).} FlexASR linear-layer has been detailed as the running example, here we focus on its role as a challenge case study: it has the largest mapping, with significant hardware data-parallelism and data-movement.

\myparagraph{HLSCNN 2D Convolution (C2D).} HLSCNN supports a 2D convolution operation that has variable kernel sizes, and supports batch computation (over filters and channels).
 
\myparagraph{HLSCNN C2D \akash{Accumulative} (C2D+).} When verifying C2D, we noted a hardware flag---unused by the 3LA mapping---for accumulating outputs onto previous results.
As the flag significantly changes data \akash{layout} correspondences, we created and verified C2D+, a mapping with the flag enabled.
\end{ignore}

\subsection{\sys by the numbers}
\label{s:evaluation-key-stats}
\hideableoutline{
\begin{itemize}
  \item Table~\ref{tab:key-stats} lists some
key statistics about each \sys step as it was applied to our case studies.
  \item (Accelerator Hex Code) Detailing the MMIO code via \lang reveals its hardware complexity.
  \item (Align step) Hardware guidance via parameter substitution and custom numeric rules enables us to synchronize all loops in the \skel.
  \item (Relate step) We use layout sketches to relate tensor data layout on the two sides of the \camappings.
  \item (Verify step) We construct a product program based on the synchronized loops. Thanks to the aligned loop nests and relational invariants in the product program, verification decomposes into multiple VC queries that are
checked using the Z3 SMT solver~\cite{demoura2008z3}, to prove equivalence of the given IR and MMIO code (verification time discussed next).
\end{itemize}}

Table~\ref{tab:key-stats} lists some
key statistics about each \sys step as it was applied to our case studies.
At first glance, the given mappings (near the top) look simple, with virtually no loops in the \aspfue (see \# loops in \aspfue in the second row).

\begin{table}
  {
  \newcommand{\most}[1]{#1}
\begin{tabular}{p{11.4em}|r r r r}
    \multicolumn{1}{r|}{Mapping}   
    & \multicolumn{1}{p{1.3em}}{\ FP} 
    & \multicolumn{1}{p{1.6em}}{FLL} 
    & \multicolumn{1}{p{1.6em}}{\raggedright C2D} 
    & \multicolumn{1}{p{1.7em}}{C2D+} \\
  \hline
  \textbf{Given code} & & & \\
  \# loops in IR code    & 2       & \most{5}  & 4     & 4 \\
  \# loops in MMIO code   & 0       & \most{1}  & 0     & 0 \\
  \hline
  \textbf{Accelerator Hex code} & & & \\
  \# application params.            & 2       & 3        & \most{6}     & \most{6} \\
  \# hardware params.              & \most{3}       & \most{3}        & 1     & 1 \\
  \# derived params.      & 2       & \most{6}        & 5     & 5  \\
  \# loops in Hex code       & 4       & \most{28}       & 12    & 12  \\
  \;\;\rotatebox[origin=c]{180}{$\Lsh$}
  max nesting depth          & 4       & 7        & \most{8}     & \most{8}  \\
  \;\;\rotatebox[origin=c]{180}{$\Lsh$}
  \# \hPfor loops            & 1   & \most{18}  & 4 & 4 \\
  \;\;\rotatebox[origin=c]{180}{$\Lsh$}
  \# \hFor loops             & 3   & \most{10}   & 8 & 8 \\
  \hline
  \textbf{Align step}  & & & \\
  \# loops in \skel          & 4        & \most{10}      & 8     & 8 \\
  \# data-movement loops     & 0        & \most{18}      & 4     & 4 \\
  \# Parameter Subst. Rules     & 2        & \most{5}       & \most{5}     & \most{5} \\
  \# Custom Numeric Rules       & 3      & \most{7}       & 3     & 3 \\
  \# rewrites to align & 12 & 29 & 27 & 27 \\
  \# loops in aligned IR code   & 4        & \most{11}      & 8     & 8 \\
  \# aligned loops         & \most{4}    & \most{10} & \most{8} & \most{8} \\
  \hline
  \textbf{Relate step}  & & & \\
  \# \dcss & 4    & 11     & \most{12}     & 8 \\
  \# data layout invariants   & 11   & \most{31}     &  10     & 5 \\
  \hline
  \textbf{Verify step}  & & & \\
  \# loops in product program & 4    & \most{29}     & 12    & 12 \\
  \# manual invariants & 11  & \most{25}     &  10    & 10 \\
  \# VC queries & 10 & 15 & 15 & 15
\end{tabular}%
}
  \caption{\sys Evaluation: Key statistics for various steps in case studies (listed in \S\ref{s:evaluation-mappings-list}).}
  \label{tab:key-stats}
  \vspace*{-6mm}
\end{table}

\myparagraph{Accelerator Hex code.}
Detailing the \aspfue using \lang (the next set of rows) reveals its hardware complexity --- \EG the code for FlexASR linear-layer (FLL) now has 28 loops, max loop nesting depth of 7, and significant hardware data-parallelism (many \hPfor loops).
Various loop bounds and dimensions are captured via application, hardware, and derived parameters in the \aspf.


\myparagraph{\emph{\stepT}.}
Each \skel (extracted from the \aspf) includes several loops 
which guide the user in transforming the \cspf for loop alignment. 
Our heuristic extraction procedure omits the many pure-data movement loops, allowing the user to focus on tensor-computation loops in the accelerator code.
Users transform the \cspf to match the \skel using our system of rewriting rules, which we customize to the accelerator using: (1) parameter substitution rules extracted from the \aspf preamble, and (2) custom numeric rewrite rules (discussed in \S\ref{s:bolt-align-rewriting}).
As an indicator of user-effort, we report the number of rewriting steps needed to align the \cspf with the \skel. 
We note that the aligned \cspf successfully matches every loop in the \skel (\# aligned loops = \# loops in \skel), with only one unaligned IR loop (in FlexASR linear-layer, FLL) that performs data initialization.

\myparagraph{\emph{\stepR}.}
Following the tactics we provide, users specify \dcss for aligned loops.
Some loops require more than one \dcs, \EG in HLSCNN 2D convolution (C2D), extra \dcss specify how HLSCNN initializes accumulators as it iterates. In contrast, C2D+ foregoes these extra \dcss, as accumulators on both sides correspond at all times.
We report the number of data layout invariants, which is high for FLL, the most complex mapping (29 loops in the product program).

\myparagraph{\emph{\stepV}.}
We automatically construct a product program based on aligned loops (note that \# loops in product program $=$ \# loops in \aspf $+$ \# loops in aligned \cspf\ $-$ \# aligned loops).
Users annotate the product program with data layout invariants and additional manual invariants in the different categories (\S\ref{s:bolt-verify})---numbers are higher for more complex mappings (\EG FLL).  
Thanks to the aligned loop nests and relational invariants,
our verifier decomposes the product program into multiple VC queries that are checked using the Z3 SMT solver~\cite{demoura2008z3}, to
prove equivalence of the given \cspf and \aspf. 


\subsection{Results for the Verify step}
\label{s:evaluation-experiments}
\hideableoutline{
\begin{itemize}
  \item Table~\ref{tab:verif-time} shows the time taken to verify the correctness of mappings using the Z3~\cite{demoura2008z3} SMT solver. Times are reported for variations of the concrete hardware/application parameters, which roughly signal the scale of the verification instance.
  \akash{[TODO]} Comparison with verification with no unrolling of inner loops.
  \item \sys successfully verified several complex mappings wrt a formal hardware semantics.
  \item Verification time grows with number loops, loop nesting, and number of data layout invariants.
  \item \akash{[TODO]} Comparison with CHC-based verification.
\end{itemize}}

\begin{table}


{%
\newcommand{\colparbox}[1]{\strut\par\parbox{14em}{\strut#1\strut}\par}
\begin{tabular}{l|l|r}
  \multirow{2}{*}{Mapping} & \multirow{2}{8em}{Parameters \qquad (HW) \quad\:\;;\: (App.)} & \multicolumn{1}{p{3em}}{Verif. time (s)} \\
  \hline
  \multirow{2}{*}{\colparbox{(FP) Add-pool \\ \textit{(4 loops, max depth 4)}}}
  & (2, 4) \quad\:\:\:\:;\: (4, 4) & 1 \\
  & (16, 16) \quad;\: (32, 64) & 361 \\
  \hline
  \colparbox{(FP) Max-pool \\ \textit{(4 loops, max depth 4)}}
  & (16, 16) \quad;\: (32, 64) & 313 \\
  \hline
  \colparbox{(FP) Mean-pool \\ \textit{(4 loops, max depth 4)}}
  & (16, 16) \quad;\: (32, 64) & 391 \\
  \hline
  \multirow{3}{*}{\colparbox{(FLL) Linear-layer \\ \textit{(29 loops, max depth 7)}}}
  & (4, 2, 8) \:\:\:\:\:;\: (4, 16, 16) & 423 \\
  & (8, 4, 8) \:\:\:\:\:;\: (8, 32, 32) & 18351 \\
  & (16, 4, 16) \:;\: (16, 64, 64) & TO \\
  \hline
  \multirow{3}{*}{\colparbox{(C2D) 2D convolution \\ \textit{(12 loops, max depth 8)}}}
  & 2 \:;\: (2, 2, 5, 5, 3, 3) & 413 \\
  & 4 \:;\: (8, 8, 12, 12, 3, 3) & TO \\
  & 8 \:;\: (8, 8, 12, 12, 3, 3) & -- \\
  \hline
  \multirow{3}{*}{\colparbox{(C2D+) 2D conv. with accum. flag \\ \textit{(12 loops, max depth 8)}}}
  & 2 \:;\: (2, 2, 5, 5, 3, 3) & 4 \\
  & 4 \:;\: (8, 8, 12, 12, 3, 3) & 7750 \\
  & 8 \:;\: (8, 8, 12, 12, 3, 3) & 20579 \\
\end{tabular}
}

  \caption{SMT verification results for mappings, with varying concrete parameter values. (TO indicates a timeout of 6 hours.)
  The mappings (in the first column) also show \# loops in the product program and maximum loop nesting depth (max depth) in parentheses.}
  \label{tab:verif-time}
  \vspace{-.5cm}
\end{table}

Table~\ref{tab:verif-time} shows the time taken to verify the correctness of mappings using the Z3~\cite{demoura2008z3} SMT solver. 
Experiments were run with a 6-hour timeout on a 16-core Intel Xeon processor (3.8 GHz CPU, 256 GiB RAM).
Times are reported for different values of the concrete hardware/application parameters, 
which roughly signal the scale of the verification instance.
The last row of a mapping reflects parameter values of real hardware; in some cases, we also report results for down-scaled versions of the hardware with smaller parameter values (for faster verification), while still preserving original loop nests and data layouts.
We discuss highlights from the results.

\myparagraph{Successful verification of mappings with respect to a formal hardware semantics.}
\sys can successfully verify complex mappings covering a variety of coarse-grained ML operations across two accelerators.
Four mappings were successfully
verified with parameter values used by the actual hardware. The remaining two were verified with downscaled parameters, still capturing key hardware features like hardware data parallelism, data movement, and complex address arithmetic.

\myparagraph{Variations across case studies.}
In general, mappings with more loops and loop nesting
take longer to verify, as they 
lead to many VC queries that are also more complex. 
Keeping loop nests constant, C2D and C2D+ show that the more data layout invariants a mapping needs, the longer the verification time.
These speed differences highlight the importance of loop nests and data layouts when verifying mappings.


\myparagraph{Handling of unaligned loops.}  We experimented with providing \skels and manual invariants for unaligned loops rather than unrolling them (the tradeoff discussed in \S\ref{s:bolt-relate-dcs}) for our mappings with unaligned loops (FLL, C2D, C2D+). Although the additional user-burden reduced verification time up to 52\% for FlexASR linear-layer (FLL), it led to similar or increased time for the others. Therefore, we unroll unaligned loops as a default, to minimize user burden.


\myparagraph{Comparison with \akash{standard verification techniques}.}
We attempted to verify our smallest equivalence checking problem, row 1 of Table~\ref{tab:verif-time}, using a state-of-the-art Constrained Horn Clause (CHC) verifier Z3/Spacer~\cite{komuravelli2016spacer,demoura2008z3}. We applied it on the same well-aligned product program, but without providing any invariants.
Even for this small problem, Z3/Spacer failed to
complete verification with a 1-day timeout.
\akash{Similarly, we tried SMT-based bounded model checking (BMC)~\cite{bmc} using Z3, and this too failed on even the smallest case studies due to extensive loop unrolling in nested loops.}
(We omitted experiments on larger problems/mappings.)



\section{Related Work}
\label{s:related-work}

\myparagraph{Verified Compilers for ML.}
Liu et al.~\cite{liu2022atl,liu2024atlverif} and Courant and Leroy~\cite{courant2021polyverif} present verified optimizing compilers from high-level IRs (tensor and affine programs, resp.) to C (or C-like) code.
Both verify compiler passes via an interactive theorem prover (Rocq). 
Neither considers verification against loop nests and data layout fixed by hardware,
nor proves correctness with respect to a hardware semantics.
CompCert~\cite{compcert} provides certified compilation of C code targeting hardware (PowerPC, ARM, x86, RISC-V), but does not support coarse-grained ML accelerator intrinsics.

\myparagraph{Translation Validation for ML compilers.}
Melchert et al.~\cite{melchert2024cascade, melchert2025cgraverif} design and verify a custom compilation flow for coarse-grained reconfigurable array (CGRA) accelerators, where they perform translation validation of individual compiler steps, using information from the compiler.
Unlike \sys, they do not address verification of black-box mappings for existing coarse-grained accelerators, such as FlexASR.
Exo~\cite{ikarashi2022exo} and Mosaic~\cite{bansal2023mosaic} are recent ML compilers with verification support. Exo optimizes affine programs via \emph{user-specified schedules} of high-level transformations; it verifies these transformations using an effect analysis.
Mosaic optimizes (sparse) tensor algebra code via \emph{auto-scheduling} of high-level transformations, which it verifies using SMT-based assertion checking (\EG of tensor bounds, sizes, etc).
Both support macro-based accelerator-specific code generation (through \emph{libraries} and \emph{external functions}, respectively), but neither verifies these \camappings with respect to a formal hardware semantics.
\sys addresses this verification gap for these compilers.
%

Pouchet et al.~\cite{pouchet2024hlsverif} verify the equivalence of High Level Synthesis (HLS) programs (written in C/C++) resulting from source-to-source transformations for optimizing ML operations (e.g., matrix multiply), 
achieving impressive performance.
However, their technique is restricted to programs where branches and loop bounds are statically interpretable---essentially it unrolls all loops---and they depend on the correctness of the HLS compiler. 
In contrast, \sys checks program equivalence in general, uses loop invariants to avoid loop unrolling, and considers a formal hardware semantics.
Verdoolaege et al.~\cite{verdoolaege2012equivalence}
 verify affine program transformations via dataflow analysis, identifying data correspondences similar to our \stepR step. Their approach requires identical operations on both sides of the equivalence problem (modulo associativity and commutativity) and does not verify correctness with respect to a formal hardware semantics.

\myparagraph{ML compiler frameworks.}
Halide~\cite{ragankelley2013halide} is a language and compiler for high-performance computing applications, including machine learning.
Cl\'ement and Cohen~\cite{clement2022halideverif} verify compilation of Halide code to an imperative language, using tensor-array relationships similar to those of our \stepR step. However,
their approach relies on compiler-provided annotations
and they do not target hardware or consider a formal hardware semantics.
Many Halide extensions define \mappings to ML accelerators~\cite{li2020heterohalide,vocke2017halidedsp,stratton2020halidedsp,korhonen2015halideaccel,gu2020halideipim}; the majority lack a formal hardware semantics for their accelerator.
3LA~\cite{huang2024_3la} automatically identifies offloading opportunities based on \mappings. TVM~\cite{chen2018tvm} with BYOC~\cite{chen2021byoc} and MLIR~\cite{lattner2021mlir} allow user-specified compilation passes that support hardware accelerators, but they have not been verified.
\akash{TensorLift~\cite{gao2026tensorlift} automatically extracts tensor-level formal semantics from RTL designs for use by MLIR compilation flows. The authors formally verify the modeling of some hardware subsystems (e.g., a PE unit), but do not verify \mappings.}
Several compilers have been developed for synthesizing hardware circuits to accelerate ML applications~\cite{eldridge2021circt,minutoli2020soda,esmaeilzadeh2021verigoodml}. Synthesis passes are trusted and do not require code generation.

\myparagraph{Automated verification of product programs.} 
Our work is inspired by many prior efforts in automated verification of product (or relational) programs.
Some representative efforts include applications in compiler validation~\cite{zaks2008crossproduct,pde}, security verification~\cite{barthe04informationflow,AlmeidaBBDE16}, and differential program analysis~\cite{os2009,LahiriMSH13}.
Prior work has highlighted the importance of \emph{loop alignment} and presented techniques to exploit alignment opportunities~\cite{terauchi05secure, barthe2011relational, zaks2008crossproduct, sousa2016cartesian, jakobs2021peqcheck,DBLP:conf/tacas/HamzaF23}, including use of CHC solvers~\cite{angelis2016relationalverif,GrigoryLpar17, unno2021constraintbased}.
\begin{ignore}
:
use of type-based analysis with self-composition~\cite{terauchi05secure}, allowing a mix of synchronized and unsynchronized loops~\cite{barthe2011relational},
exploiting structurally equivalent program fragments~\cite{zaks2008crossproduct}, lockstep execution of loops~\cite{sousa2016cartesian}, synchronizing Horn clause rules in CHC solving~\cite{angelis2016relationalverif,GrigoryLpar17}, automatic inference of synchronization and relational invariants~\cite{unno2021constraintbased} (and many others).
\end{ignore}
The Align 
and Relate steps of \sys
are inspired by these prior efforts. In addition, we have proposed the \skel and the \dcs to guide loop alignment and annotation of relational invariants, respectively.
We note that none of the prior work has considered problem instances with complexity similar to our case studies, in terms of number of loops, deeply nested loops, and multidimensional and dynamic data layouts. 

\section{Conclusions and Future Directions}
\label{s:conclusion}
\begin{ignore} 
  In this work we introduce \sys, the \emph{first} (to our knowledge) framework for verifying \mappings for ML accelerators.
  Although proving equivalence between the two sides of a \mapping is 
  challenging due to differences in loop structure and complex data layout, \sys provides specialized components to overcome these challenges:
  \begin{itemize}
    \item \lang for explicating the loop structure and data layout of the accelerator hardware
    \item \emph{\skeleton} for guiding the transformation of \cspfs to align with \aspfs
    \item \emph{{\dcsfull}es} for constructing relational data layout invariants. 
  \end{itemize}
  These components are used as part of \sys's four steps --- Explicate, Transform, Relate, and Verify --- to construct a \emph{well-aligned} tensor product program and 
  verify the equivalence 
  by using an SMT solver.
  We have developed a prototype of \sys/\lang and used it to formally verify complex \mappings for two state-of-the-art accelerators.
  (We plan to make the prototype and the case studies publicly available.) \\

  \noindent
  We would like to improve \sys in the following directions in future work:
  \begin{itemize}
  \item Invariant synthesis:
  To fully automate Relate and Verify, we are exploring \emph{synthesis} of the invariants used by our verifier.
  We are particularly keen on data layout invariants, which are highly structured and where a grammar is extractable from the tensor/buffer reads and writes in the well-aligned product program.

  \item Automated \cspf transformation:
  We aim to take cues from prior work, which already performs some significant automatic transformations of tensor programs. The hardware guidance we provide (from Explicate and \skel) can reduce some exploration effort, but we may need to extend previous methods to work efficiently with custom numerics.

  \end{itemize}

\end{ignore} 

In this work we introduce \sys, 
the first framework for verifying \mappings for coarse-grained ML accelerator intrinsics, with respect to a formal hardware semantics. We use program equivalence techniques based on product programs, 
where a \skel additionally guides the user in transforming the \cspf to align with the loop nests in the \aspf, and a \dcs provides a sketch for a user to annotate relational data layout invariants on aligned loops. Our evaluation shows that \sys can successfully verify many complex mappings for recent open-source 
ML accelerators, including those with 
multiple deeply-nested loops and dynamic data layouts.
In future work, we would like to automatically generate some key invariants using automated synthesis techniques, and develop a comprehensive set of rewriting transformations via equality saturation.


\bibliographystyle{plain}
\bibliography{sources}


\end{document}